\documentclass[preprint,nofootinbib,longbibliography]{revtex4-1}
\usepackage[utf8]{inputenc}
\usepackage{amsmath}
\usepackage{hyperref}
\usepackage{amsfonts}
\usepackage{amssymb}
\usepackage{graphicx}
\usepackage{latexsym}
\usepackage{color}
\usepackage{booktabs}
\usepackage{dcolumn}
\usepackage{epsfig}
\usepackage{subfigure}
\usepackage{float}
\usepackage{multirow}
\usepackage{ulem}
\usepackage{xcolor}

\begin{document}

\title{Holographic Phase Transitions in $(2+1)$-dimensional black hole spacetimes in NMG}

 \author{E. Abdalla}\email{eabdalla@if.usp.br}
 \affiliation{Instituto de F\'isica, Universidade de S\~ao Paulo, CEP 05315-970, S\~ao Paulo, SP, Brazil}
 \author{Jeferson de Oliveira}\email{jeferson@gravitacao.org}
 \affiliation{Instituto de F\'i­sica, Universidade Federal de Mato Grosso, CEP 78060-900, Cuiab\'a, MT, Brazil}
 \author{A. B. Pavan}\email{alan@unifei.edu.br}
 \affiliation{Instituto de F\'isica e Qu\'imica, Universidade Federal de Itajub\'a, CEP 37500-903, Itajub\'a, MG, Brazil}
 \author{C. E. Pellicer}\email{carlos.pellicer@ufrn.br}
 \affiliation{Escola de Ci\^encias e Tecnologia , Universidade Federal do Rio Grande do Norte, Caixa Postal 1524, 59072-970, Natal, RN, Brazil}
\date{\today}

%------------------------------------------------------------%
\begin{abstract}

In this work we aim at the question of holographic phase transitions in two dimensional systems with Lifshitz scaling. We
consider the gravity side candidate for a dual description as the black hole solution of New Massive Gravity (NMG) with Lifshitz scaling. We discuss the effects due to the Lifshitz scaling in the AGGH (Ayon-Beato-Garbarz-Giribet-Hassa\"\i ne) solution
in comparison with the BTZ (Ba\~nados-Teitelboim-Zanelli) black hole. Likewise, we compute the order parameter and it indicates a second order phase transition in a $(1+1)$ dimension Lifshitz boundary.

\end{abstract}
%------------------------------------------------------------%

\maketitle

%------------------------------------------------------------%

\section{Introduction}\label{introduction}

The Anti-de Sitter/conformal field theories correspondence (AdS/CFT)
turned out to be a very useful tool to map the physics of a quantum
field theory at strong coupling in $D-1$ dimensions to a classical
gravity theory in $D$ dimensions whose spatial infinity is
isometric to the AdS spacetime \cite{maldacena}\cite{witten}\cite{wittenklebanov}.

Much attention has been given to the extensions of
such a correspondence regarding the study of condensed matter
systems defined at the AdS boundary, such as superconductivity and superfluidity \cite{3Hs}\cite{binpapa},
non-fermi liquids \cite{liu1} and strange metals \cite{hartnollpolch}.

In order to study condensed matter systems described by
non-relativistic theories, we need solutions to the gravity side which
exhibit the so-called Lifshitz scaling
\cite{son}\cite{kachru}\cite{balasumb}. Recently, a black hole
solution with such a symmetry was found in the
context of New Massive Gravity (NMG) in three dimensions
\cite{chile}. Therefore,temperature can be added to the holographic
description resulting in a non-relativistic field theory at finite temperature at the boundary.

Such a black hole solution is stable under scalar and spinor perturbations
\cite{berthapellicerjef} and from the point of view of the AdS/CFT
correspondence the IR limit is a dual description of an integrable model system
given by the Korteweg-de Vries (KdV) equation \cite{kdv}.

This paper is organized as follows. In section~\ref{sec-background} a brief
review of the Lifshitz black hole in three dimensions is presented. In section~\ref{sec-eom}
the equations of motion for the matter fields in the bulk are derived and
analyzed in the probe limit. In section~\ref{sec-phasetrans}, using a semi
analytical analysis, we obtain the phase transition in the Lifshitz
boundary and the critical electric field where it occurs.
In section~\ref{sec-numeric} we numerically solve the equations of motion and derive the
order parameters. Finally in section~\ref{sec-remarks} we conclude and discuss some open questions.

%------------------------------------------------------------%
\section{The gravity background and the matter fields}\label{sec-background}

The NMG is a three dimensional theory of a spin-$2$ field \cite{townsend}
equivalent to the unitary Pauli-Fierz theory \cite{fierzpauli} at linearized level.
Moreover, a version of NMG with a non-vanishing cosmological constant
was considered in \cite{chile} and the corresponding action reads
\begin{equation}\label{actionNMG}
S_{g}=\frac{1}{16\pi G}\int d^{3}x\sqrt{-g}\left(R-2\Lambda-K\right),
\end{equation}
where $R$ is the Ricci scalar, $\Lambda$ is the three-dimensional cosmological constant and $K$
encodes the higher curvature terms,
\begin{equation}\label{hcurvature}
K=\frac{1}{\bar{m}^{2}}\left[R_{\mu\nu}R^{\mu\nu}-\frac{3}{8}R^{2}\right],
\end{equation}
with $\bar{m}$ being the graviton mass in three dimensions. Looking for black hole solutions in the equations of motion of the action (\ref{actionNMG}), is assumed a line element
\begin{equation}\label{lineelement}
ds^{2}=-\frac{r^{2z}}{l^{2z}}f(r)dt^{2}+\frac{l^2}{r^2f(r)}dr^{2}+r^{2}d\phi^{2}\quad ,
\end{equation}
where $z$ is the dynamic exponent which determine two different black hole solutions.

The first one, when $z=3$, is a black hole solution named the AGGH black hole. It exhibits the anisotropic scale invariance $t\rightarrow\lambda^{z}t$, $\vec{x}\rightarrow\lambda\vec{x}$. The line element that describes the geometry for this black hole~\cite{chile} is given by
\begin{equation}\label{metric_z3}
ds^{2}=-\frac{r^6}{l^6}f(r)dt^{2}+\frac{l^2}{r^2f(r)}dr^{2}+r^{2}d\phi^{2}\quad ,
\end{equation}
where
\begin{equation}\label{efe}
f(r)=\left(1-\frac{r_{+}^{2}}{r^{2}}\right)\quad ,
\end{equation}
with $r_{+}=l\sqrt{M}$ denoting the event horizon location, $M$ is related to the black hole mass and
$l=\sqrt{-13/32\Lambda}$ is the AdS radius. The spacetime represented by such a solution has a light-like
singularity at $r=0$. The spatial infinity ($r\rightarrow \infty$) has some properties similar to the AdS spacetime \cite{berthapellicerjef}.

The second solution, for $z=1$, is the well-known BTZ (Ba\~nados-Teitelboim-Zanelli) black hole solution \cite{BTZ}
\begin{equation}\label{metric_z1}
ds^{2}=-\frac{r^2}{l^2}f(r)dt^{2}+\frac{l^2}{r^2f(r)}dr^{2}+r^{2}d\phi^{2}\quad ,
\end{equation}
where $f(r)$ is given by (\ref{efe}), the event horizon is located at $r_{+}=l\sqrt{M}$ covering the
singularity at $r=0$ and in the limit $r\rightarrow \infty$ the solution is AdS-like. Thus, the NMG allows us to study the relativistic case $z=1$ and the non-relativistic case $z=3$ in the same setup. The theory provides a scenario to observe the role of Lifshitz symmetry in the formation of the holographic phase transitions in comparison to the relativistic case $z=1$.

Thus, we take as a background, the geometry given by the
three-dimensional black holes of NMG. These solutions have all the
main features needed in order to apply the gauge/gravity holographic
prescription for phase transitions: there is an AdS-like spatial
infinity and a regular event horizon, whose presence is necessary for
the condensation of a charged scalar field.

The action describing a charged scalar field $\Psi$ coupled to gravity
and to the electromagnetic field in three dimensions can be written as
\begin{equation}\label{action_fields}
S_{f}=\int d^{3}x\sqrt{-g}\left[-\frac{1}{4}F_{\mu\nu}F^{\mu\nu}-|\nabla\Psi-iqA\Psi|^{2}-m^{2}|\Psi|^2\right],
\end{equation}
where $F_{\mu\nu}=\nabla_{\mu} A_{\nu}-\nabla_{\mu}A_{\nu}$, $q$ is the scalar field charge and $m$ its
mass. Here, we consider the scalar and gauge fields in the probe limit. This means
that the fields do not backreact on the geometry. Thus, in order to describe the phase transition it
is enough to consider the equations of motion for the matter fields evolving in the fixed background of
the metrics (\ref{metric_z3}) or (\ref{metric_z1}). If we perform the field rescaling $\Psi\rightarrow
\Psi/q$, $A_{\mu}\rightarrow A_{\mu}/q$, the probe limit can be understood as the limit
$q\rightarrow\infty$. Since in this limit the action of matter fields behaves as $q^{-2}$ they
decouple from gravity, whose action behaves as $q^{0}$.

Finally, an important question arises when we work in $(2+1)$- dimensional gravity.
 Clement \cite{clement}  pointed out that rotating charged three-dimensional black holes present a
logarithmic divergence in the mass and angular momentum. In such a case,  the author introduced a
Chern-Simons term in the Einstein-Maxwell action in order to heal those divergencies. However, Ba\~nados
et al \cite{BTZ} have shown that for static charged BTZ such a divergence in the mass can be handled.
Thus, we expect that not to be a problem in our cases. In fact, in a quick inspection we observe that
the introduction of a Chern-Simons term in the action \ref{action_fields} modifies the coupling between
Abelian and scalar fields. The effect of this term in the phase transition of the system described
on the boundary  will be addressed  in a future work.

%------------------------------------------------------------%
\section{Equations of motion and symmetries}\label{sec-eom}

In this section, we are going to present the equations of motion for the fields $\Psi$ and $A_{\mu}$
in the probe limit, showing the role of the scaling symmetries of these fields in the equations.

The equations of motion for $\Psi$ and $A_{\mu}$ are, respectively,
\begin{eqnarray}\label{full_psi}
&&\nabla^{\mu}\nabla_{\mu}\Psi+2iqA_{\mu}g^{\mu\nu}\nabla_{\nu}\Psi+iq g^{\mu\nu}\Psi\nabla_{\nu}A_{\mu}
-q^{2}g^{\mu\nu}A_{\mu}A_{\nu}\Psi -m^{2}\Psi=0\quad ,\\
\label{full_phi}
&&\nabla^{\mu}F_{\mu\nu}=2q^{2}A_{\nu}\Psi^{2}\quad ,
\end{eqnarray}
where we have taken $\Psi$ to be real, without loss of generality. For
our purposes, it is enough to consider the fields depending only on the radial coordinate $r$ as
\begin{equation}\label{ansatz_psi_phi}
\Psi=\Psi(r),\hspace{0.3cm} A=\phi(r)dt .
\end{equation}
 Then eqs. (\ref{full_psi})-(\ref{full_phi}) reduce to
\begin{eqnarray}\label{psi_u}
\Psi^{\prime\prime}(u)+\left[\frac{f^\prime(u)}{f(u)}-\frac{z}{u}\right]\Psi^\prime(u)+
\left[\frac{q^{2}l^{2(z+1)}u^{2(z-1)}}{f(u)^{2}r_{+}^{2z}}\phi(u)^{2}-
\frac{\alpha^{2}}{f(u)u^{2}}\right]\Psi(u)&=&0\quad ,\\
\label{phi_u}
\phi^{\prime\prime}(u)+\frac{z}{u}\phi^\prime(u)-\frac{2q^{2}l^{2}\Psi(u)^{2}}{f(u)u^{2}}\phi(u)&=&0\quad ,
\end{eqnarray}
where $\alpha= m l$ and $z=1$, $z=3$ correspond to the dynamical exponents for BTZ and AGGH cases,
respectively. The coordinate $u=r_{+}/r$ is the new radial coordinate which maps the event horizon and
the boundary to the interval $[1,0]$ and ${}^ \prime$ represents the derivative with respect to $u$ coordinate.

The above system of differential equations exhibits a very useful
scaling symmetry for the fields $\Psi(u)$ and $\phi(u)$. If we perform
the redefinitions
\begin{equation}\label{symmetry}
\Psi(u)\rightarrow\frac{1}{ql}\hat{\Psi}(u),\hspace{0.3cm}\phi(u)\rightarrow\frac{2\pi
T_H}{q}\hat{\phi}(u),
\end{equation}
where $T_H$ is the Hawking temperature,
\begin{equation}\label{temp}
T_H=\frac{1}{2\pi}\frac{r_{+}^{z}}{l^{z+1}}\quad ,
\end{equation}
the equations of motion (\ref{psi_u})-(\ref{phi_u}) can be cast in
the dimensionless form
\begin{eqnarray}\label{phi_semT}
\hat{\phi}^{\prime\prime}(u)+\frac{z}{u}\hat{\phi}^\prime(u)-\frac{2\hat{\Psi}(u)^{2}}{f(u)u^{2}}\hat{\phi}&=&0\quad ,\\
\label{psi_semT}
\hat{\Psi}^{\prime\prime}(u)+\left[\frac{f^\prime(u)}{f(u)}-\frac{z}{u}\right]\hat{\Psi}^\prime(u)+
\left[\frac{u^{2(z-1)}}{f(u)^{2}}\hat{\phi}^{2}(u)-
\frac{\alpha^{2}}{f(u)u^{2}}\right]\hat{\Psi}(u)&=&0\quad ,
\end{eqnarray}
without an explicit dependence on the black hole temperature. As we
will see in detail in the next section, the phase transition will be
governed by the value of the electric field due the scalar field
condensate in the neighborhood of the event horizon.

Furthermore, it is worth mentioning that the equations of motion (\ref{phi_semT}) (\ref{psi_semT}) are invariant
under the anisotropic scale invariance $t\rightarrow \lambda^{z}$, $r\rightarrow \lambda^{-1}r$ if
\begin{equation}\label{simetria_escala_campos}
\phi\rightarrow\lambda^{-z}\phi, \hspace{0.3cm} \Psi\rightarrow \Psi\quad ,
\end{equation}
and the Hawking temperature scales as
\begin{equation}\label{temp_escala}
T_{H}\rightarrow \lambda^{-z}T_{H}.
\end{equation}

Looking into the solutions (\ref{phi_z}) we see that
\begin{equation}\label{rho_mu_escala}
\rho\rightarrow \lambda^{-z}\rho, \hspace{0.3cm}\mu\rightarrow \lambda^{-z}\mu\quad .
\end{equation}

Thus, comparing (\ref{rho_mu_escala}) and (\ref{temp_escala}) we can build up the variable $T_{H}/\mu$ as playing
the role of our temperature parameter in order to eliminate the scale factor $\lambda$ from the description. Therefore, we set
\begin{equation}\label{temperatura_3hs}
\hat{T}=\frac{T_{H}}{\mu}\quad ,
\end{equation}
implying that the critical temperature $T_{c}\propto \mu$.

%------------------------------------------------------------%
\section{The phase transition and the critical electric field}\label{sec-phasetrans}

In this section we obtain an approximate expression for the dual operators $\langle\mathcal{O}_{1}\rangle$ and
$\langle\mathcal{O}_{2}\rangle$ in terms of the asymptotic behavior of
the solutions for the fields $\Psi$ and $\phi$ following the standard
AdS/CFT correspondence \cite{3Hs} \cite{kanno}. In summary the process consists
in finding the leading order solutions in the region near the black
hole event horizon $u=1$ and in the spatial infinity $u=0$, then match
the two sets of solutions at an intermediate radius $u=u_{0}$ requiring continuity of the functions $\Psi$ and $\phi$. The validity of
these approximations and matching conditions are discussed in \cite{Kim:2013oba}.

The result is an approximate expression for the phase transition
and consequently the critical value of the order parameter which
controls the charged scalar field condensation. Therefore, we
will be able to see explicitly the condensate dependence on the Lifshitz exponent $z$.

\subsection{Solutions at spatial infinity $u\rightarrow 0$}

The solutions for the fields $\phi(u)$ and $\Psi(u)$\footnote{We omit the \emph{hat} notation.} from the
eqs. (\ref{phi_semT}) and (\ref{psi_semT}) in the spatial infinity are given by
\begin{eqnarray}\label{phi_z}
\phi(u) &=& \left\{
\begin{array}{rl}
\rho + \mu \ln{u} & \text{if} \hspace{0.2cm}z = 1,\\ \\
\rho + \mu \frac{u^{1-z}}{1-z} & \text{if} \hspace{0.2cm} z \neq 1. \\
\end{array} \right.\\
\nonumber\\
\label{psi_z}
\Psi(u)&=&C_{1}u^{\Delta_{+}}+C_{2}u^{\Delta_{-}},
\end{eqnarray}
with
\begin{equation}\label{deltas}
\Delta_{\pm}=\frac{(z+1)}{2}\pm\frac{1}{2}\sqrt{(1+z)^{2}+4\alpha^{2}}\quad ,\hspace{0.3cm}
\text{and}\hspace{0.2cm} \alpha=m l\quad ,
\end{equation}
where $\rho$ is identify as the charge density of the dual field theory living at
$u=0$ and $\mu$ its chemical potential. The factors $C_1$ and $C_2$ will be identified as the expectation value $\langle\mathcal{O}_{1}\rangle$ and $\langle\mathcal{O}_{2}\rangle$ of the dual operators in the AdS-like border.
As we see in the expression (\ref{phi_z}) the leading term is $\ln(u)$ (for $z=1$), and $u^{1-z}$ (for $z\not=1$), its coefficient is interpreted as a
chemical potential and the subleading term as the charge density.

An interesting effect of the Lifshitz symmetry in the evolution of the scalar field can be observed inspecting the
conformal dimension of its dual operator in eq.\eqref{deltas}. Beside the evident fact that the Lifshitz exponent $z$
increases the conformal dimension, when $z\neq0$ a new BF bound to the mass of the scalar field is obtained
\begin{equation}
\label{nbfbound}
\alpha_{BFL}^2=-\frac{(1+z)^{2}}{4}.
\end{equation}
For $z=3$ the BF-Lifshitz bound $\alpha_{BFL}^2=-4$ is smaller than the traditional BF bound $\alpha_{BF}^2=-1$
for $(2+1)$ dimensions. Thus, the presence of the Lifshitz symmetry expands the range of mass of the scalar field affecting
the conformal dimension of the operator living on the boundary.

It is important to stress that to obtain the asymptotic fields presented in the eqs. (\ref{phi_z}) and (\ref{psi_z})
we had to impose restrictions on the values of the scalar field mass. In order to obtain the asymptotic solution
to $\phi$ we had to impose that $\Psi\to 0$ faster than $u$ near to the boundary $(u\to0)$ resulting in a condition that
must be satisfied, that is, $\Delta_{\pm} > 1$ . Because of this restriction the permitted range of the mass of the scalar field changes.

\begin{figure}[!htbp]
\centering
\includegraphics[height=0.45\textwidth,angle=-90]{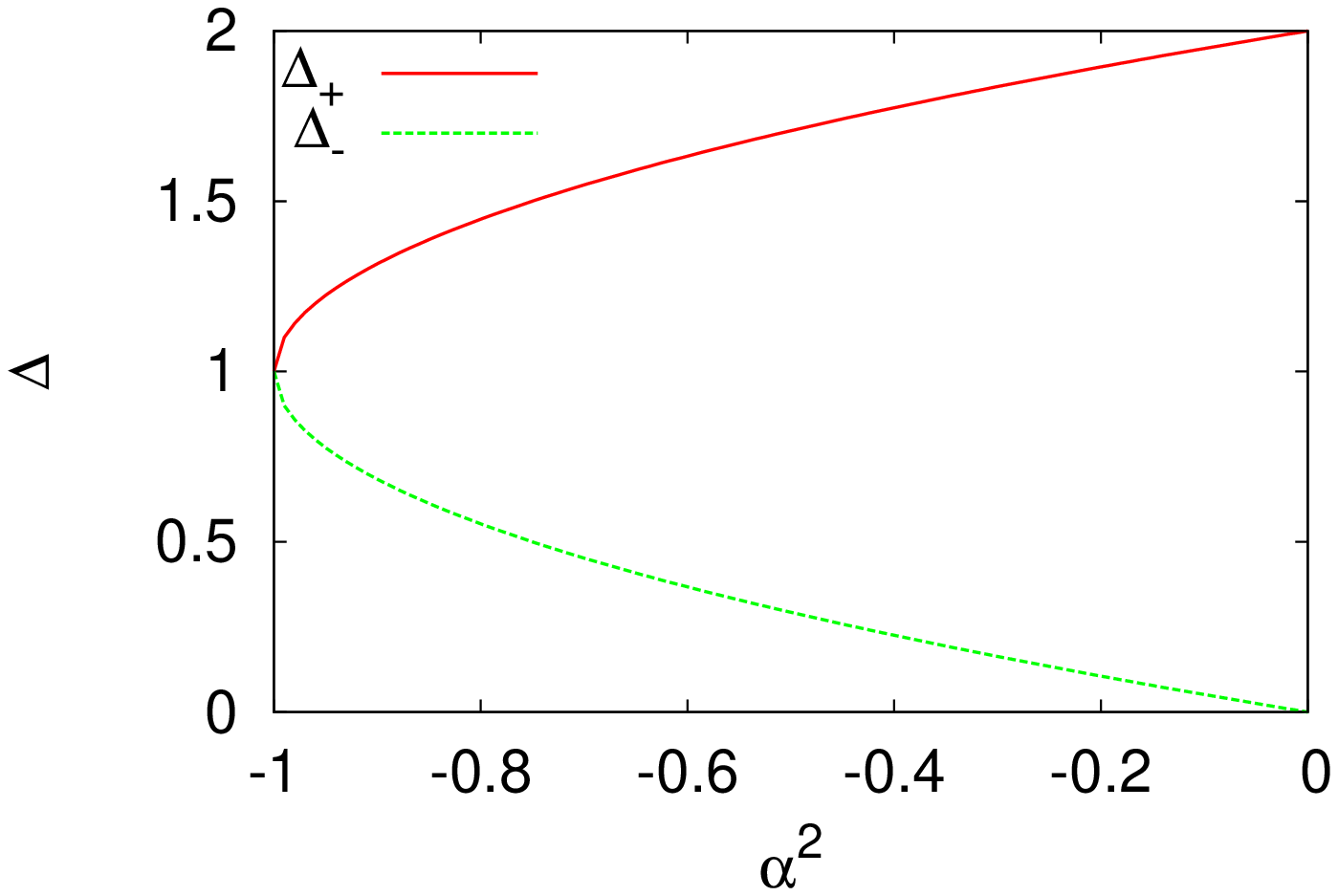}
\includegraphics[height=0.45\textwidth,angle=-90]{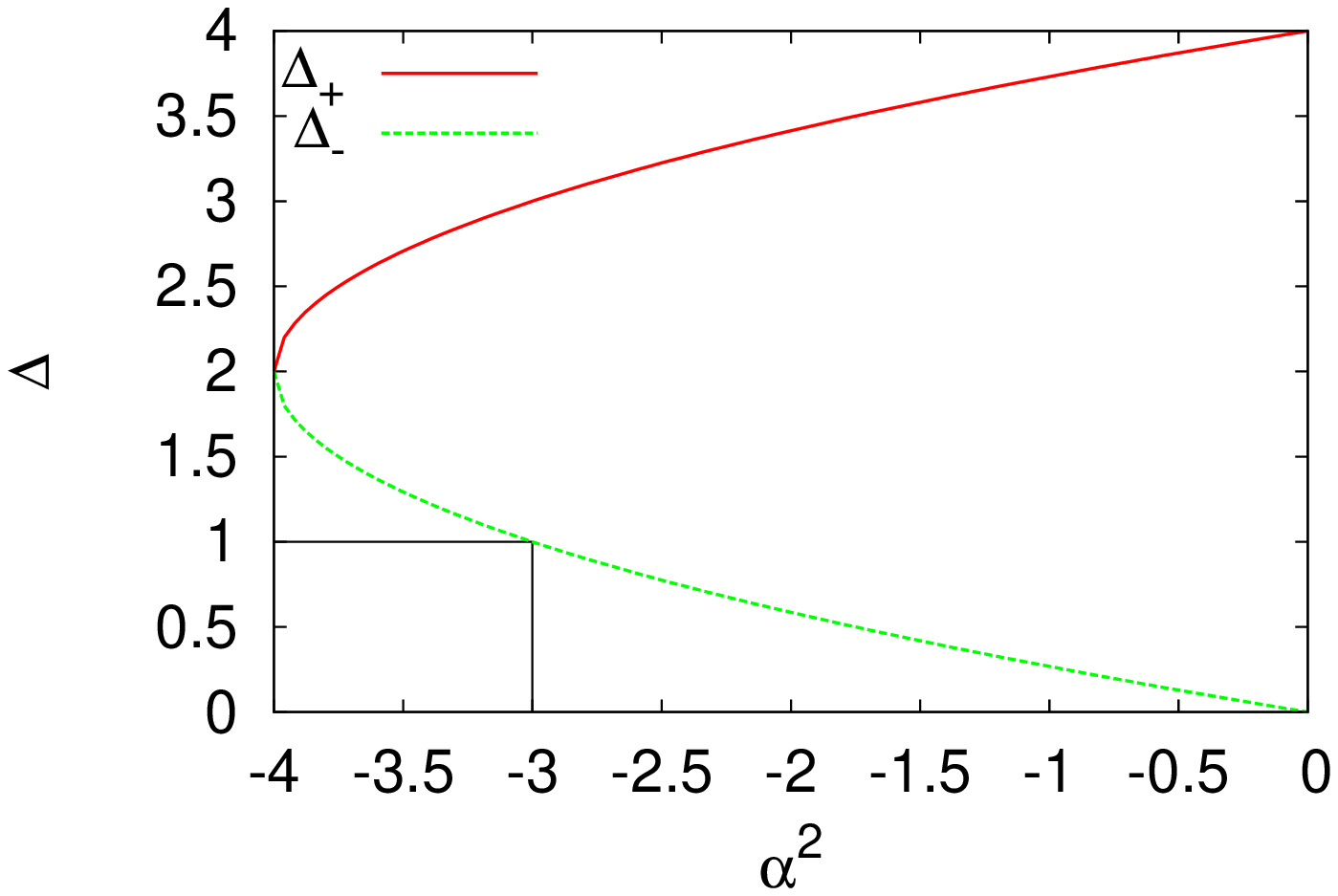}
\caption{Limits of mass for $\Delta_{\pm}$ for BTZ case (left) and in AGGH (right)}
\label{delta_limit}
\end{figure}

In figure~\ref{delta_limit} we show these ranges according to the conformal dimension $\Delta$ for BTZ and AGGH
black holes. For the BTZ black hole such a restriction excludes $\Delta_{-}$ as a possible conformal dimension for all
range of mass while for $\Delta_+$ the permitted range will be \linebreak $-1\leq\alpha^2\le 0$. For the AGGH
black hole, the conformal dimension $\Delta_{-}$ will have the range restricted to $-4\leq\alpha^2\leq-3$ while
for $\Delta_{+}$ the range will be $-4\leq\alpha^2\le 0$. This limit is consistent with \cite{kachru}.
We exclude positive values of $\alpha^2$ for both cases
because $\Delta_{-}<0$ and $\Psi$ would diverge as $u$ tends to $0$.

\subsection{Solutions at the event horizon $u\rightarrow 1$}

In order to have a finite electric potential at event horizon  we must
impose
\begin{equation}\label{phi_u1}
\phi(1)=0,
\end{equation}
and eq. (\ref{psi_semT}) implies
\begin{equation}\label{psi_u1}
\Psi^\prime(1)=-\frac{\alpha^{2}}{2}\Psi(1).
\end{equation}
We expand the fields $\Psi$ and $\phi$  near the
event horizon $u=1$ as
\begin{eqnarray}\label{taylor_psi}
\Psi(u)&=&\Psi(1)+(u-1)\Psi^\prime(1)+\frac{1}{2}(u-1)^{2}\Psi^{\prime\prime}(1)+\cdots,\\
\label{taylor_phi}
\phi(u)&=&\phi(1)+(u-1)\phi^\prime(1)+\frac{1}{2}(u-1)^{2}\phi^{\prime\prime}(1)+\cdots.
\end{eqnarray}
Expanding the equation of motion for $\Psi$ (\ref{psi_semT}) near
$u=1$ an substituing $\Psi^{\prime\prime}(1)$ in the expansion (\ref{taylor_psi}),
we have
\begin{equation}\label{taylor_psi2}
\Psi(u)=\Psi(1)-\frac{1}{2}\alpha^{2}(u-1)\Psi(1)-\left[\frac{1}{8}\alpha^{2}(3-z)+
\frac{1}{16}\alpha^{4}-\frac{1}{2}\phi^\prime(1)^{2}\right](u-1)^{2}\Psi(1)^{2}.
\end{equation}

The same procedure for the electric potential $\phi$ leads to
\begin{equation}\label{taylor_phi2}
\phi(u)=\left[(u-1)-\frac{z+\Psi(1)^{2}}{2}(u-1)^{2}\right]\phi^\prime(1),
\end{equation}
where in the above two expression we have imposed the regularity
conditions at the event horizon (\ref{phi_u1}) and (\ref{psi_u1}).

\subsection{Matching the solutions at $u=u_{0}$}

Having the solutions (\ref{phi_z}) and (\ref{psi_z}) at spatial infinity and (\ref{taylor_phi2}) and (\ref{taylor_psi2}) near
the event horizon, we can
connect these two sets of solutions smoothly at a radius $u=u_{0}$,
which can be arbitrary without changing the main features of the phase
transition. We begin by the BTZ black hole $(z=1)$, whose connection relations at $u=u_{0}$ are
\begin{eqnarray}\label{match_z1a}
\rho +
\mu\ln{u_{0}}&=&\left[(1-u_{0})+\frac{1}{2}\left(1+a^{2}\right)(1-u_{0})^{2}\right]b\quad ,\\\label{match_z1b}
\frac{\mu}{u_{0}}&=&-\left[1+(1+a^{2})(1-u_{0})\right]b\quad ,\\ \label{match_z1c}
C_{1}u_{0}^{\Delta_{+}}&=&\left\{1-\frac{\alpha^{2}}{2}(u_{0}-1)+\frac{1}{2}(u_{0}-1)^{2}\left[\frac{\alpha^{2}}{2}+
\frac{\alpha^{4}}{8}-\frac{b^{2}}{8}\right]\right\}a\quad ,\\\label{match_z1d}
\Delta_{+}C_{1}u_{0}^{\Delta_{+}-1}&=&\left\{-\frac{\alpha^{2}}{2}+\left[\frac{\alpha^{2}}{2}+\frac{\alpha^{4}}{8}-
\frac{b^{2}}{8}\right](u_{0}-1)\right\}a\quad ,
\end{eqnarray}
where we have defined $a\equiv\Psi(1)$ and $b\equiv -\phi^\prime(1)$ and
taken $C_2=0$ in order to find $C_1$. On the other
hand, if we take $C_1=0$ we get $C_2$.

Solving eq. (\ref{match_z1b}) for $a^{2}$,
\begin{equation}\label{order_parameter}
a^{2}=-\frac{\mu}{b(1-u_{0})u_{0}}\left[1+\frac{\left(2-u_{0}\right)u_{0}b}{\mu}\right]\quad .
\end{equation}
For the charged scalar field to condense near the event horizon,
we see that $b/\mu$ must be negative, since $a$ is assumed to be real, therefore $a^2>0$.

From  eqs. (\ref{match_z1c}) and (\ref{match_z1d}) we find
\begin{equation}\label{C1}
C_{1}=\Gamma_{+}\left(\frac{b_{c}}{b}\right)^{\frac{1}{2}}\left(1-\frac{b}{b_{c}}\right)^{\frac{1}{2}},
\end{equation}
where the critical value for $b$ , denoted by $b_{c}$, is given by
\begin{equation}\label{bc}
b_{c}=\frac{|\mu|}{\left(2-u_{0}\right)u_{0}}\quad,
\end{equation}
and
\begin{equation}\label{Gammaz1}
\Gamma_{+}=\frac{1}{2\ u_{0}^{\Delta_{+}-1}}\left[\frac{4+\alpha^{2}(1-u_{0})}
{2u_{0}+\Delta_{+}(1-u_{0})}\right]\left[\frac{2-u_{0}}{1-u_{0}}\right]^{\frac{1}{2}}\quad .
\end{equation}

 Using the AdS/CFT dictionary,  eq.~\eqref{C1} can be read off as the expectation value $\langle\mathcal{O}_{1}\rangle$
of the operator dual to the charged scalar field $\Psi$,
\begin{equation}\label{O1z1}
\langle\mathcal{O}_{1}\rangle^{\frac{1}{\Delta_{+}}}=\Gamma_{+}^{\frac{1}{\Delta_{+}}}
\left(\frac{b_{c}}{b}\right)^{\frac{1}{2\Delta_{+}}}\left(1-\frac{b}{b_{c}}\right)^{\frac{1}{2\Delta_{+}}}\quad .
\end{equation}

As expected, $\langle\mathcal{O}_{1}\rangle$  is zero at the critical value
of the electric field $b=b_c$, the charged scalar field
condensates and, of course, the phase transition occurs for
$b<b_{c}$. The exponent $1/2$ shows us the general behavior of mean field
theory for a second order phase transition. For the AGGH black
hole ($z=3$) the same qualitative behaviour is observed and the structure of a  mean field theory is preserved at the boundary.

Now, considering $C_1=0$ and following the same steps for $C_2$,  we find that the expectation value $\langle\mathcal{O}_{2}\rangle$ for the BTZ black hole
is given by
\begin{equation}\label{O2z1}
\langle\mathcal{O}_{2}\rangle^{\frac{1}{\Delta_{-}}}=\Gamma_{-}^{\frac{1}{\Delta_{-}}}
\left(\frac{b_{c}}{b}\right)^{\frac{1}{2\Delta_{-}}} \left(1-\frac{b}{b_{c}}\right)^{\frac{1}{2\Delta_{-}}}.
\end{equation}

Thereafter, the same procedure was performed for the AGGH black hole $(z=3)$. We just list the
results for the two operators,
\begin{eqnarray}\label{O1z3}
\langle\mathcal{O}_{1}\rangle^{\frac{1}{\Delta_{+}}}_{z >
  1}&=&\Gamma_{+,z > 1}^{\frac{1}{\Delta_{+}}}\left(\frac{b_{c}}{b}\right)^{\frac{1}{2\Delta_{+}}}
\left(1-\frac{b}{b_{c}}\right)^{\frac{1}{2\Delta_{+}}}\quad ,\\
\label{O2z3}
\langle\mathcal{O}_{2}\rangle^{\frac{1}{\Delta_{-}}}_{z >
  1}&=&\Gamma_{-,z > 1}^{\frac{1}{\Delta_{-}}}\left(\frac{b_{c}}{b}\right)^{\frac{1}{2\Delta_{-}}}
\left(1-\frac{b}{b_{c}}\right)^{\frac{1}{2\Delta_{-}}}\quad ,
\end{eqnarray}
where
\begin{equation}\label{Gammaz3}
\Gamma_{+,z > 1}=\frac{1}{2\ u_{0}^{\Delta_{+}-1}}\left[\frac{4+\alpha^{2}(1-u_{0})}{2u_{0}+\Delta_{+} (1-u_{0})}\right]\left[\frac{1+z(1-u_{0})}{1-u_{0}}\right]^{1/2}\quad ,
\end{equation}
\begin{equation}\label{Gamma2z3}
\Gamma_{-,z > 1}=\Gamma_{+,z > 1}\quad ,
\end{equation}
if we exchange $\Delta_{+}$ for $\Delta_{-}$.
The critical value of the electric field is
\begin{equation}\label{bcz3}
b_{c}=\frac{|\mu|}{u_{0}^{z}\left[1+z(1-u_{0})\right]}\; , \hspace{0.3cm}\mu<0\quad .
\end{equation}

\newpage
\section{Numerical results for the phase transition}\label{sec-numeric}

In this section we numerically solve eqs. (\ref{phi_semT}) and (\ref{psi_semT}). They form a system
of coupled second order ordinary differential equations, which can be solved using fourth order
Runge-Kutta method. We input the boundary conditions at the event horizon $u=1$ ($r=r_+$) and find
the values of $\Psi(u)$ and $\phi(u)$ on a grid $u = 1 - i*\Delta u$ with $i\in (0,1,\ldots,N-1)$,
$\Delta u = 1.0/N$ and $N=1000$.

Equations (\ref{phi_u1}) and (\ref{psi_u1}) fix two conditions, but at this point
we still do not have a condition for $\Psi(1) = \Psi_+$ and $\phi^\prime(1) = E_+$. So, for
each pair $(\Psi_+,E_+)$ we integrate eqs. (\ref{phi_semT}) and (\ref{psi_semT}) to look
for a convenient
behaviour. As $u \to 0$ ($r \to \infty$), $\Psi(u)$ behaves as eq. (\ref{psi_z}) and $\phi(u)$
behaves as
eq. (\ref{phi_z})
which are linear on the
parameters, so we can use the least square method to calculate the asymptotic behaviour. With this
procedure, we have a map
\begin{equation}\label{sm_map}
(\Psi_+,E_+) \to (C_1,C_2, \rho, \mu) \ .
\end{equation}
We are interested in the cases where $C_2=0$, in which we define $\langle O_1 \rangle =C_1$. Similarly,
for $C_1=0$ we define $\langle O_2 \rangle =C_2$.
Using the shooting method, we search for pairs of boundary values for $(\Psi_+,E_+)$ mapped to such
conditions.
We vary $E_+$ from $0$ to $15$ in $1500$ steps for $z=1$ and from $0$ to $35$ in $3500$ steps for $z=3$.
\footnote{We started varying $E_+$ from $0$ to $10$ and later we changed in order to see at least
five curves.}
For each $E_+$, we vary $\Psi_+$ from $0$ to $10$ in $10000$ steps.
\footnote{We noticed that we needed smaller steps in $\Psi_+$ for the $<O_1>$ and
$<O_2>$ curves to be smooth.}

For $z=1$, keeping $E_+$ fixed, we assume that $C_1$ and $C_2$ are smooth functions of $\Psi_+$ when using
the map (\ref{sm_map}).

Thus, whenever $C_2$ changes sign, we add a point to the graph of $\langle O_1 \rangle$
as function of $E_+$ and, whenever $C_1$ changes sign, we add a point to the graph of $\langle O_2 \rangle$
as a function of $E_+$.

\begin{figure}[!htbp]
\centering
\includegraphics[height=0.45\textwidth,angle=-90]{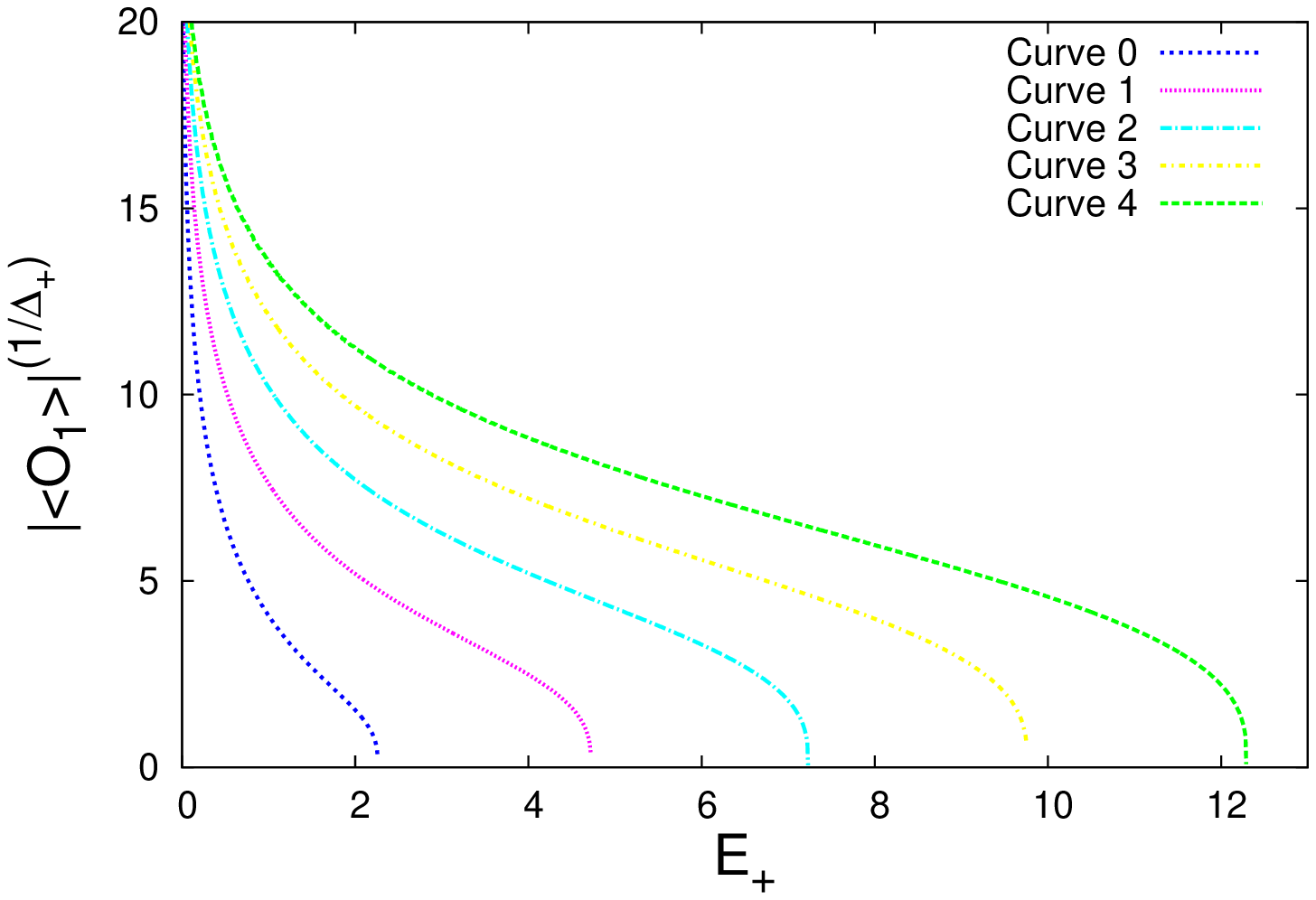}
\includegraphics[height=0.45\textwidth,angle=-90]{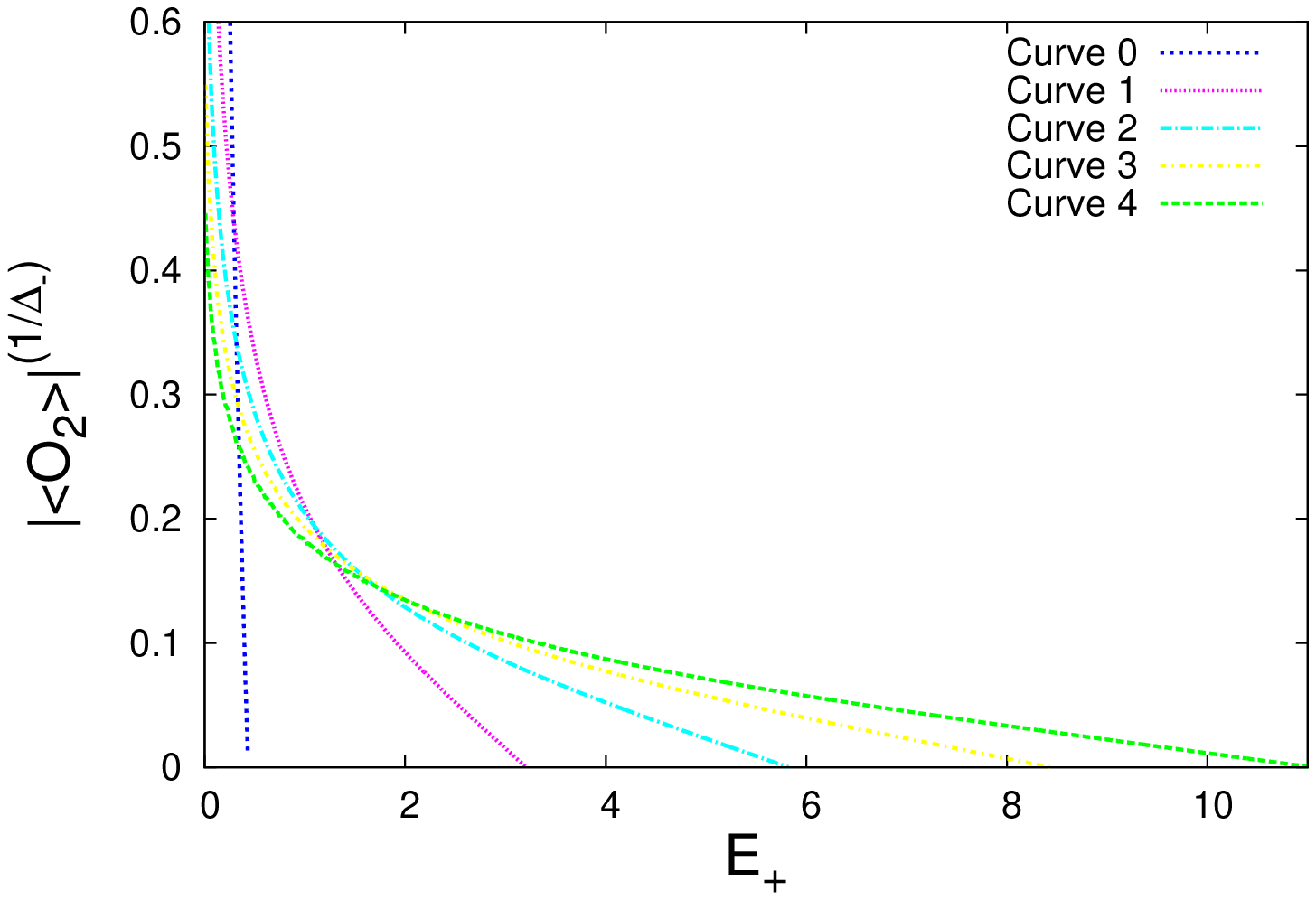}
\caption{First five curves for $z=1$ and $\alpha^2 = -0.75$.}
\label{graf_db_psi_12_E_z_1}
\end{figure}

\begin{figure}[!htbp]
\centering
\includegraphics[height=0.45\textwidth,angle=-90]{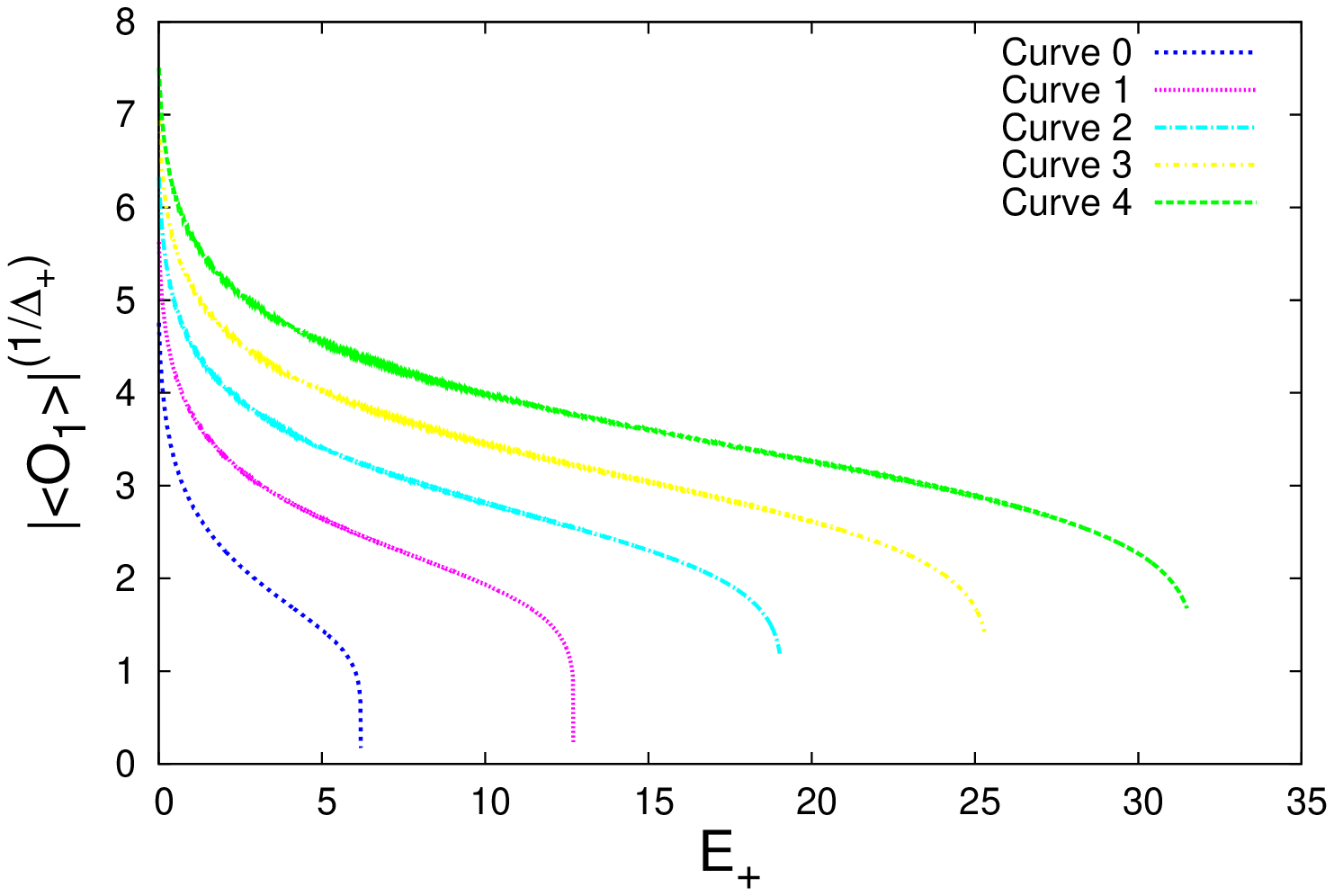}
\includegraphics[height=0.45\textwidth,angle=-90]{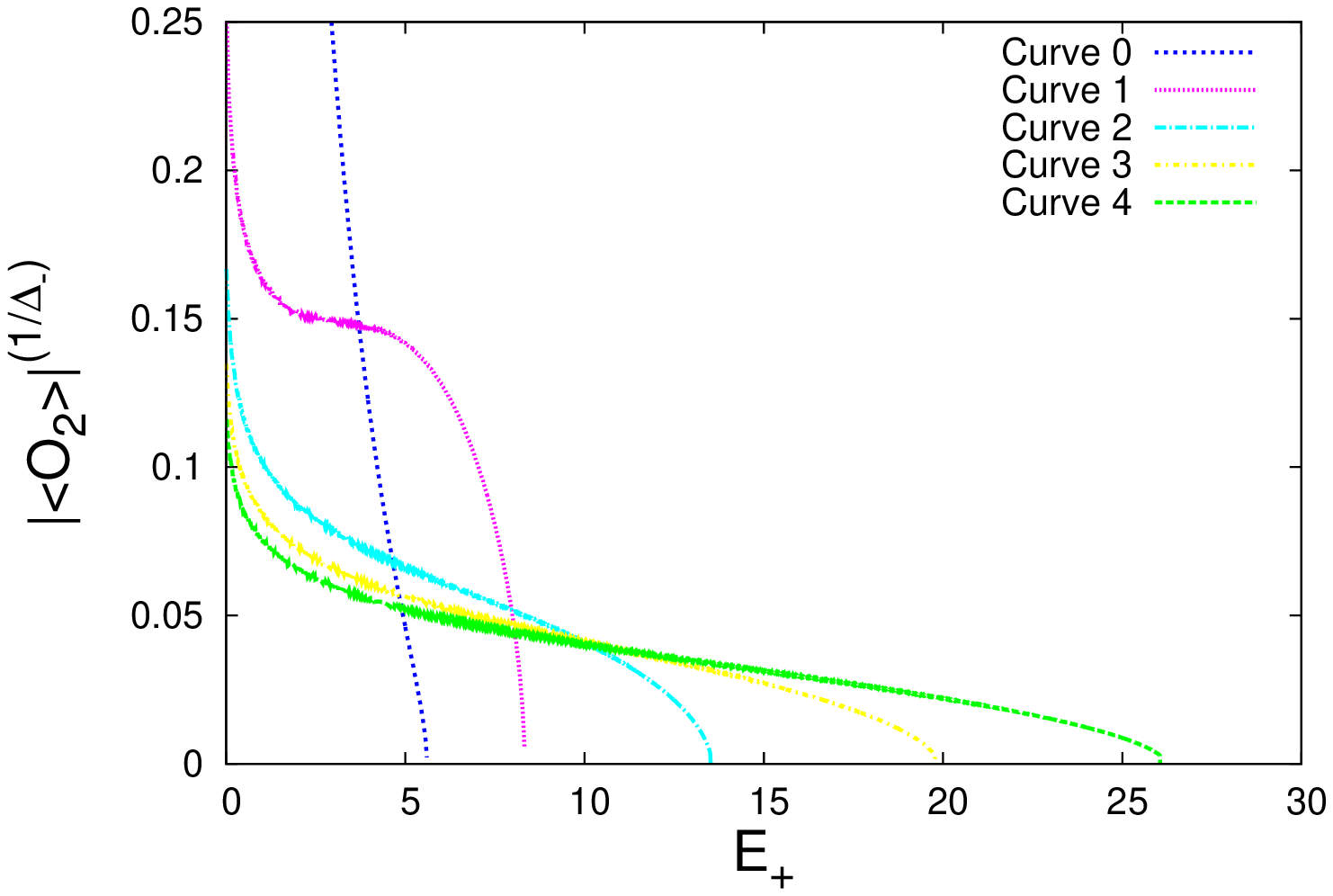}
\caption{First five curves for $z=3$ and $\alpha^2 = -2.75$.}
\label{graf_db_psi_12_E_z_3}
\end{figure}

We notice that for $z=1$, if we plot all the data our algorithm generates, we see
several scattered points and in the middle of these points we can see smooth curves that
go to zero as we raise $E_+$. If the first occurrences of sign change are isolated as one vary $\Psi_+$,
the isolated points correspond to the smooth curves observed. These curves can also be labelled by the number
of times $\Psi(u)$ changes sign, allowing us to identify the curve labelled as 0 as the fundamental mode and the others as excited modes.
In \cite{apop} the role of the excited modes in the phase transition is explained.
We choose to plot the five first occurrences of sign change in figure~\ref{graf_db_psi_12_E_z_1}.
For $z=3$, no scattered points appear in our data. Even so, we choose to plot the five first occurrences of sign change in figure~\ref{graf_db_psi_12_E_z_3}.
One interesting property is that for $\langle O_1 \rangle$, none of the smooth curves crosses another, while for
$\langle O_2 \rangle$, each curve crosses every other.

\begin{figure}[!htbp]
\centering
\includegraphics[height=0.45\textwidth,angle=-90]{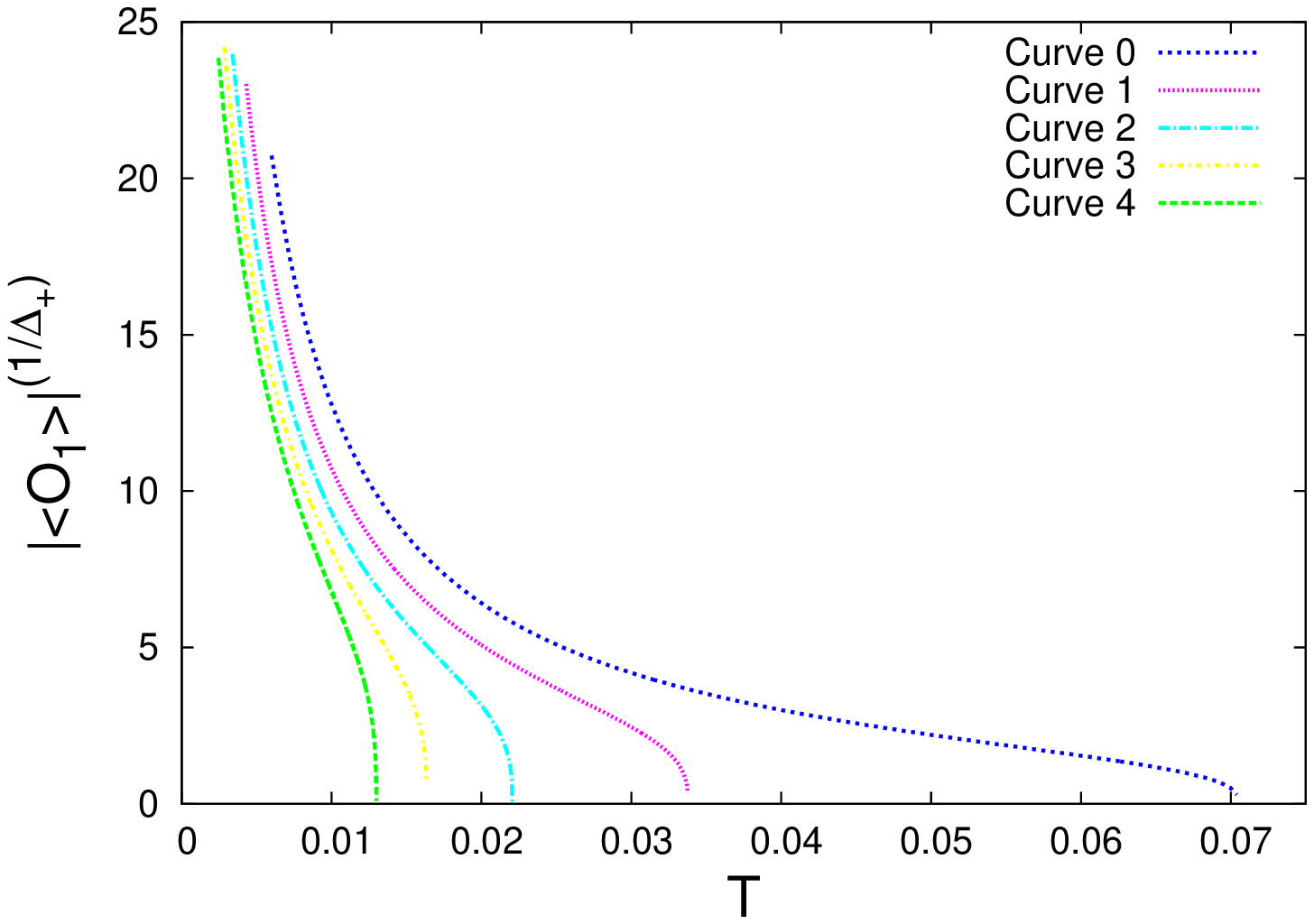}
\includegraphics[height=0.45\textwidth,angle=-90]{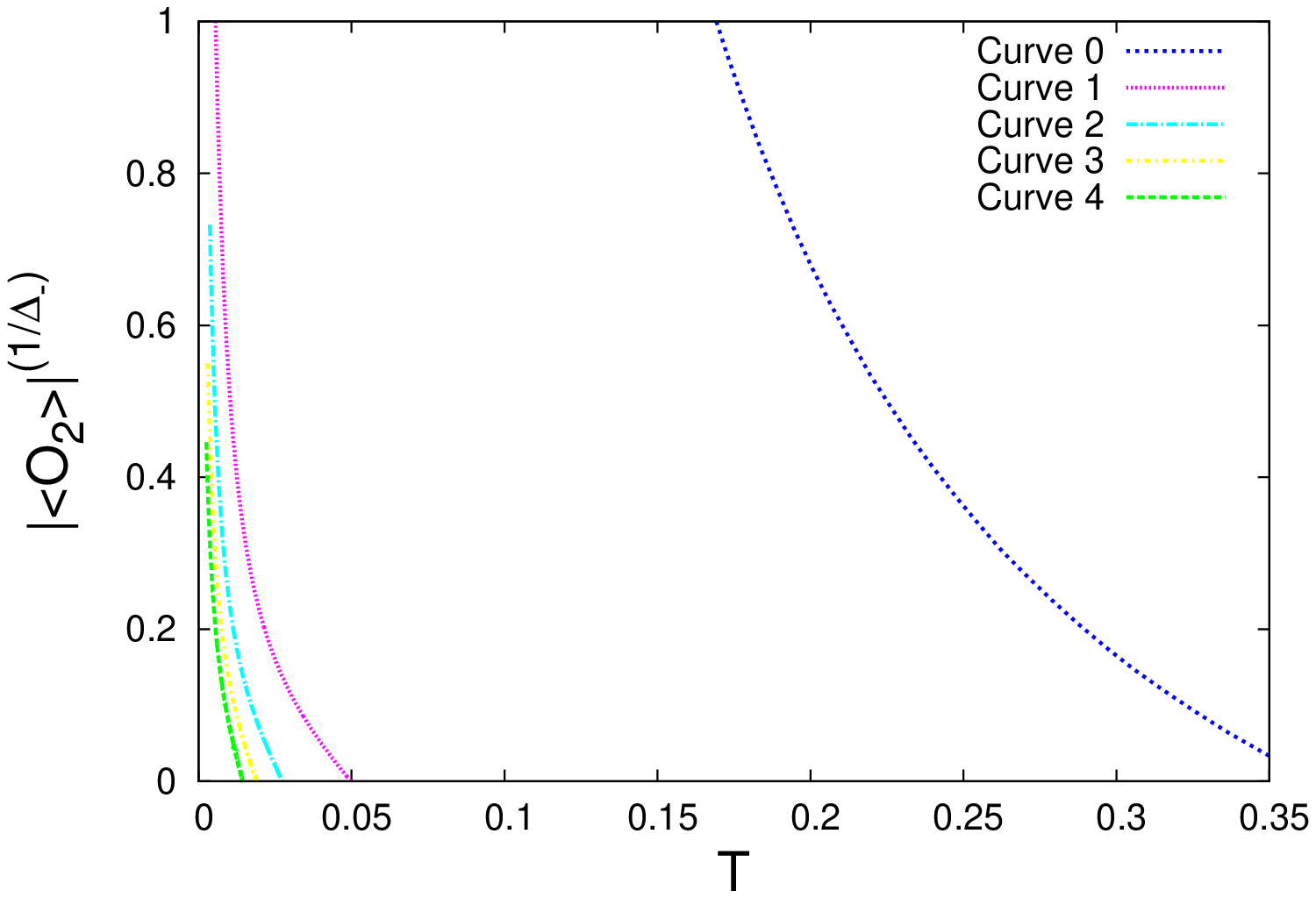}
\caption{First five curves for $z=1$ and $\alpha^2 = -0.75$.}
\label{graf_db_psi_12_T_z_1}
\end{figure}

\begin{figure}[!htbp]
\centering
\includegraphics[height=0.45\textwidth,angle=-90]{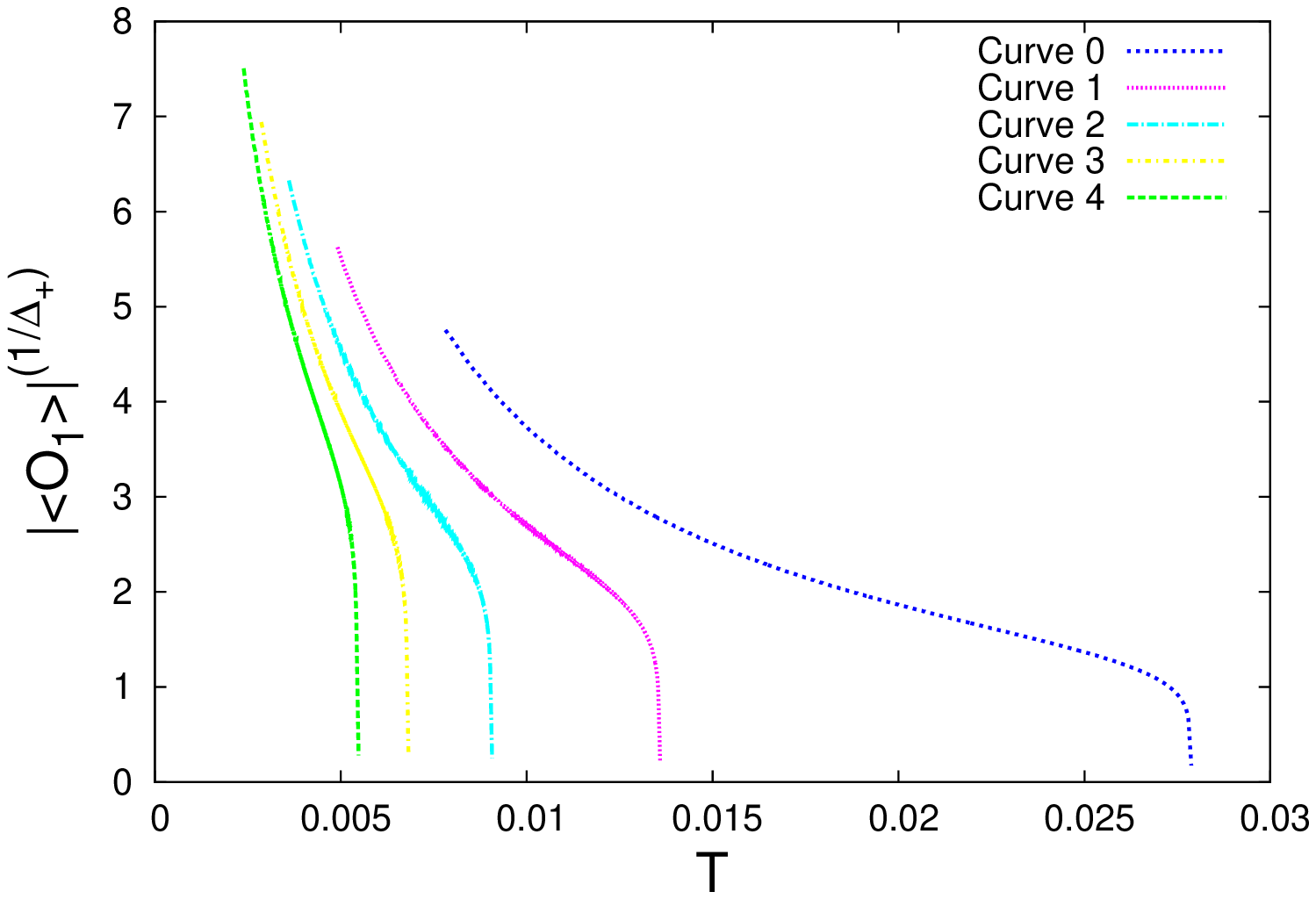}
\includegraphics[height=0.45\textwidth,angle=-90]{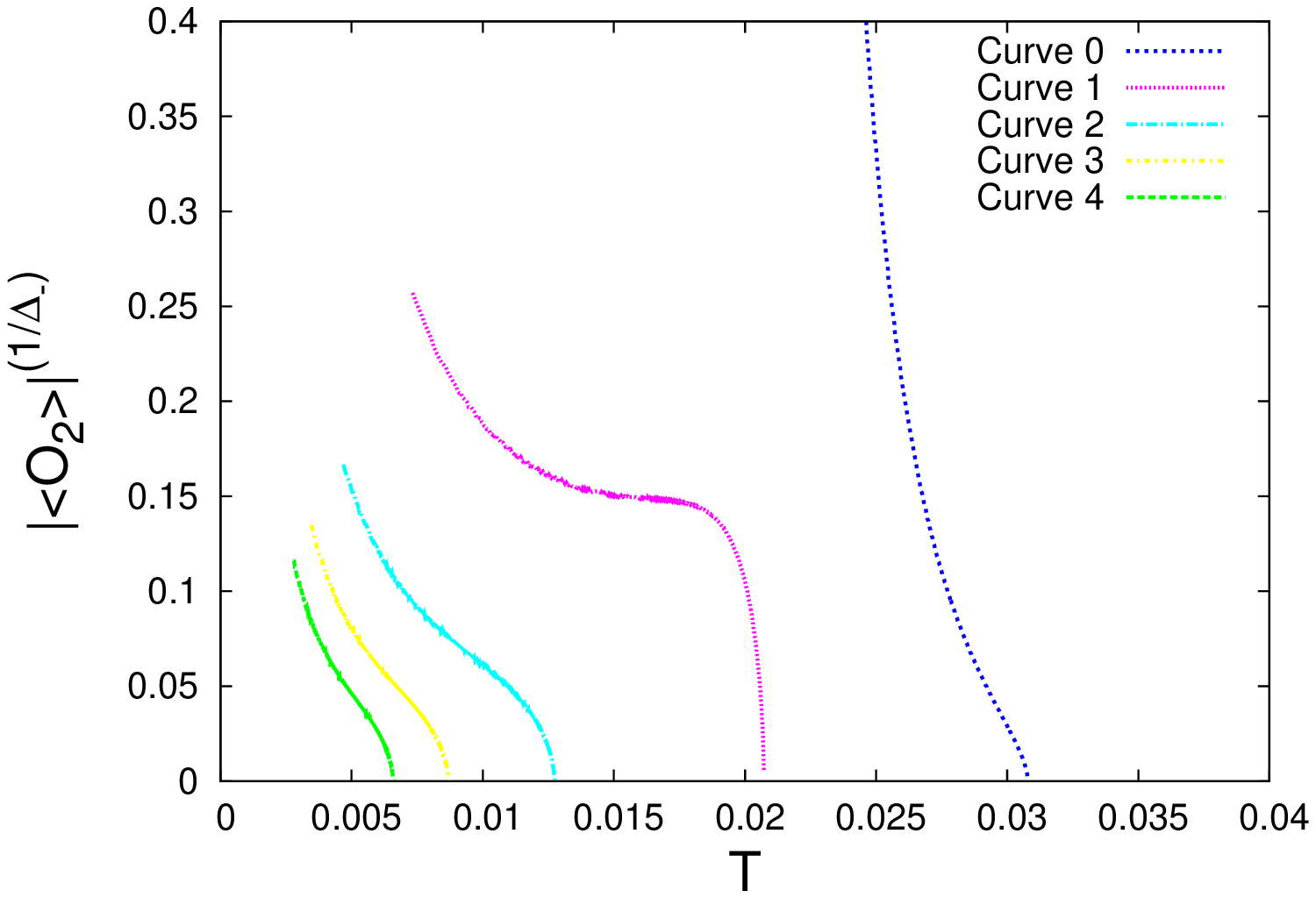}
\caption{First five curves for $z=3$ and $\alpha^2 = -2.75$.}
\label{graf_db_psi_12_T_z_3}
\end{figure}

We can also see the dependence of $\langle O_1 \rangle$ and $\langle O_2 \rangle$ on the variable $T = 1/(2\pi\mu)$ defined in eq. (\ref{temperatura_3hs}) with the arbitrary choice $r_+=l=1$. This dependence is plotted in figure~\ref{graf_db_psi_12_T_z_1} for $z=1$ and figure~\ref{graf_db_psi_12_T_z_3} for $z=3$. If the incoherent points not shown in figures~\ref{graf_db_psi_12_E_z_1} and~\ref{graf_db_psi_12_E_z_3} are plotted, one see that they do not appear to be incoherent anymore, as they are now concentrated in a region of low temperatures. The curves shown behave as an order parameter of a phase transition.

\begin{figure}[!h]
\centering
\includegraphics[height=0.45\textwidth,angle=-90]{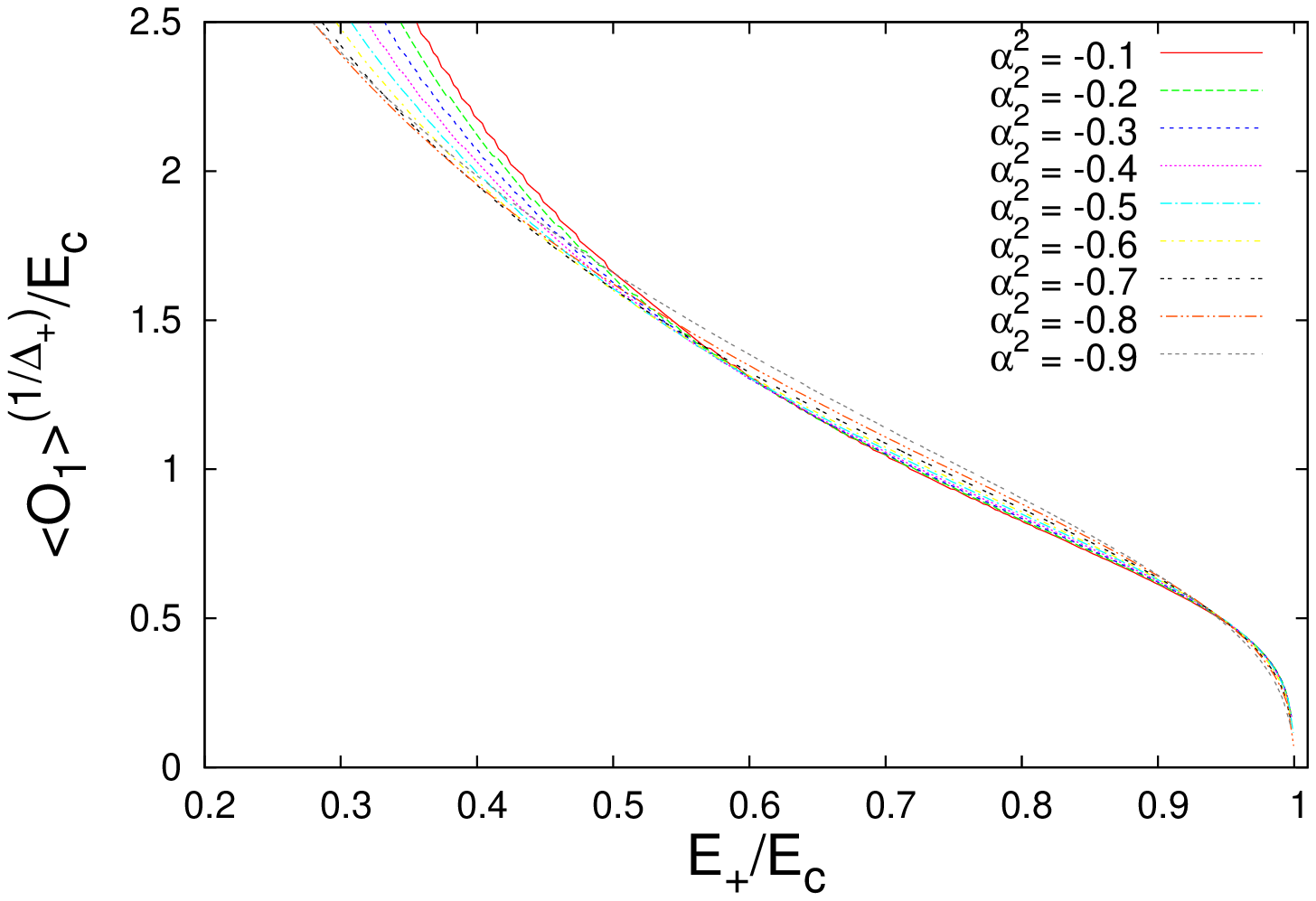}
\includegraphics[height=0.45\textwidth,angle=-90]{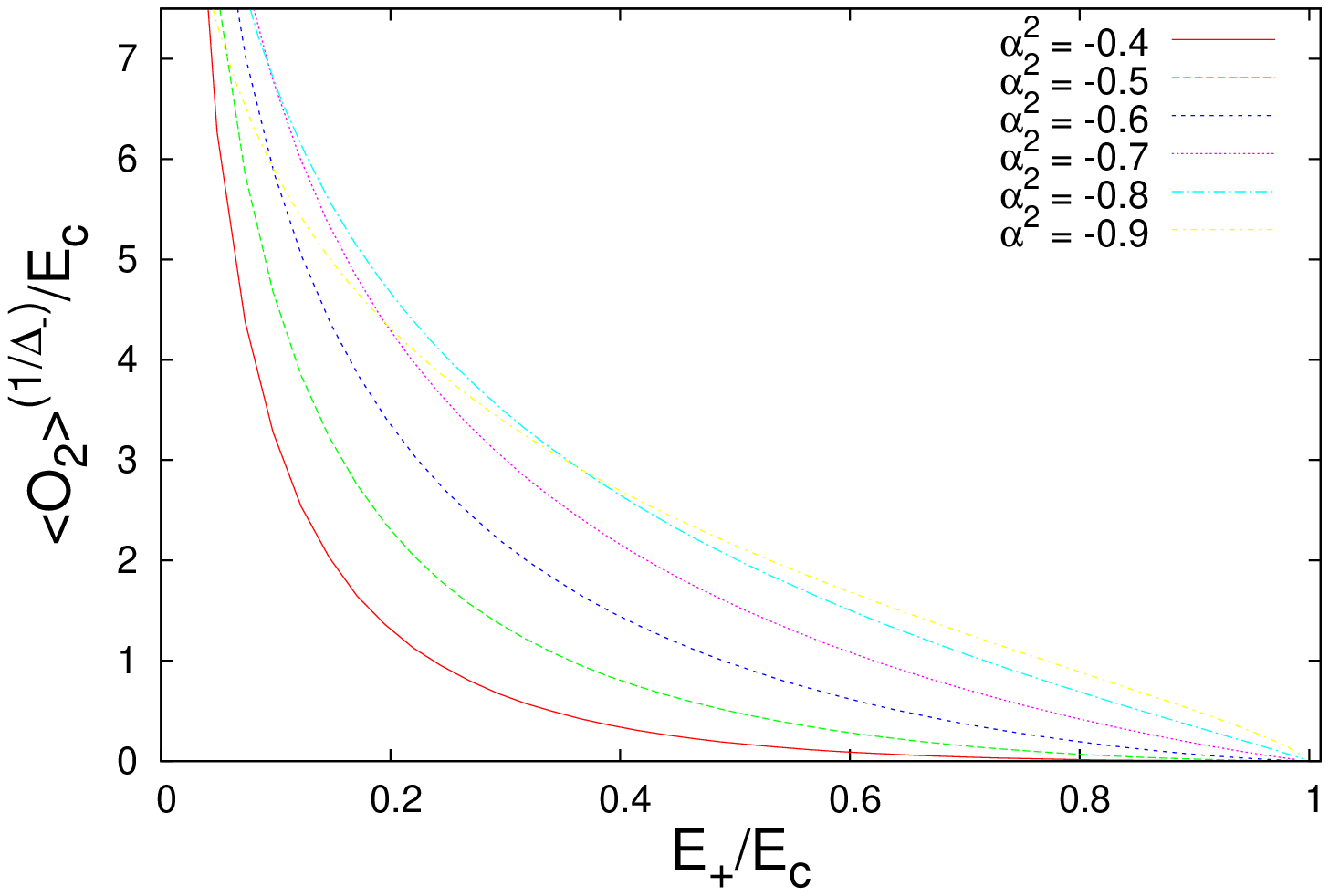}
\caption{Order parameters dependent on $E_+$ for $z=1$.}
\label{graf_cn_psi_12_E_z_1}
\end{figure}

\begin{figure}[!h]
\centering
\includegraphics[height=0.45\textwidth,angle=-90]{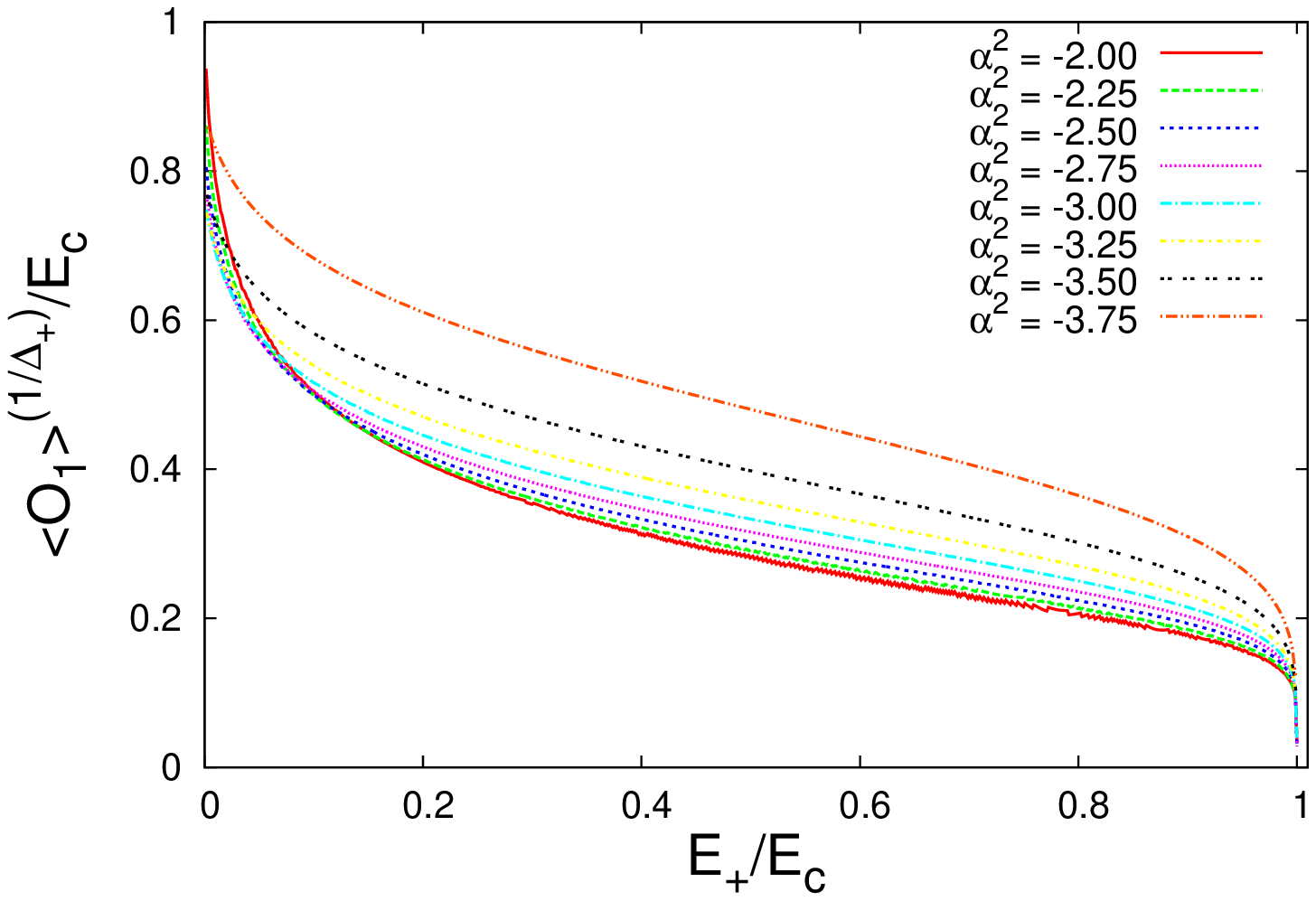}
\includegraphics[height=0.45\textwidth,angle=-90]{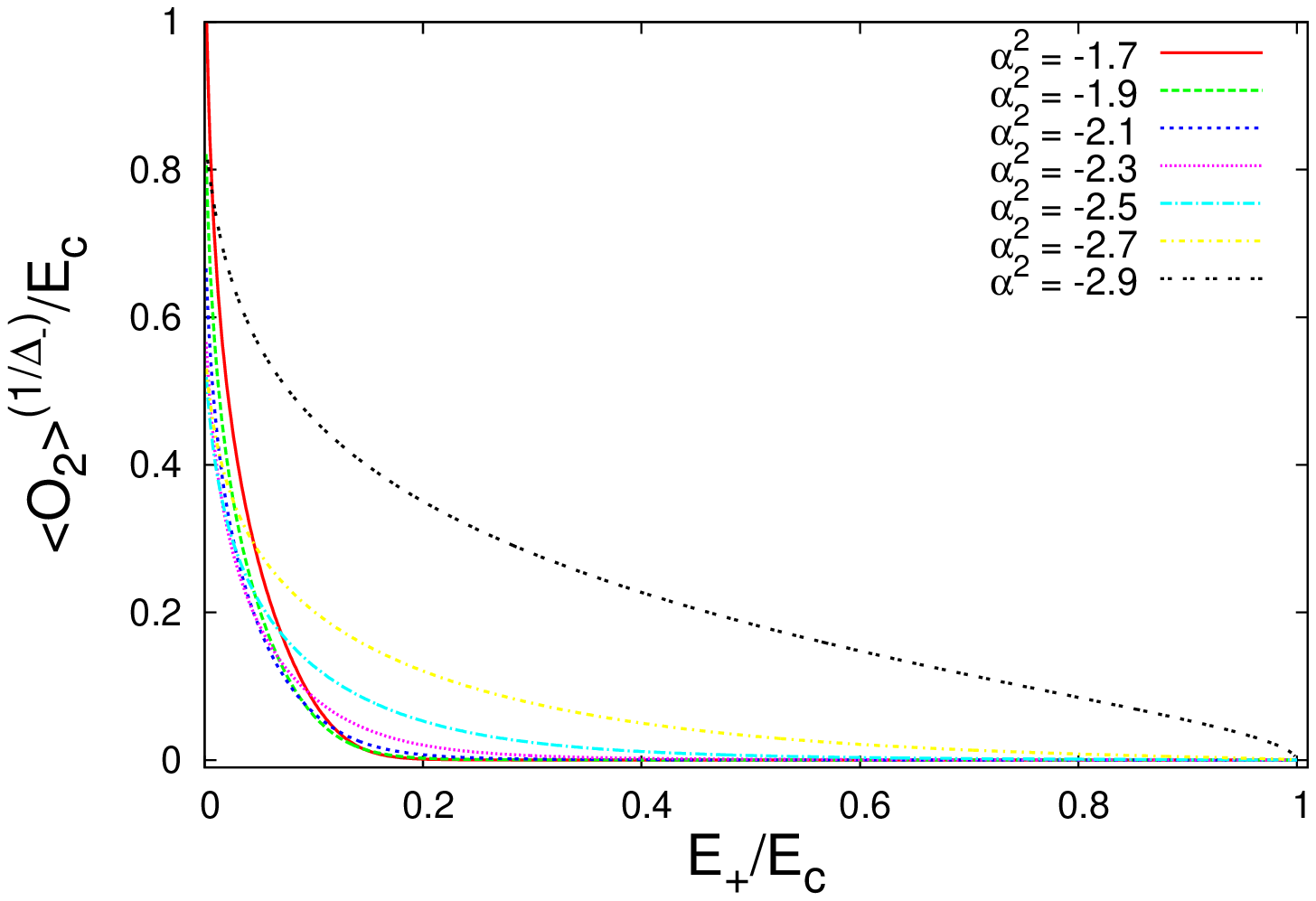}
\caption{Order parameters dependent on $E_+$ for $z=3$.}
\label{graf_cn_psi_12_E_z_3}
\end{figure}

\begin{figure}[!h]
\centering
\includegraphics[height=0.45\textwidth,angle=-90]{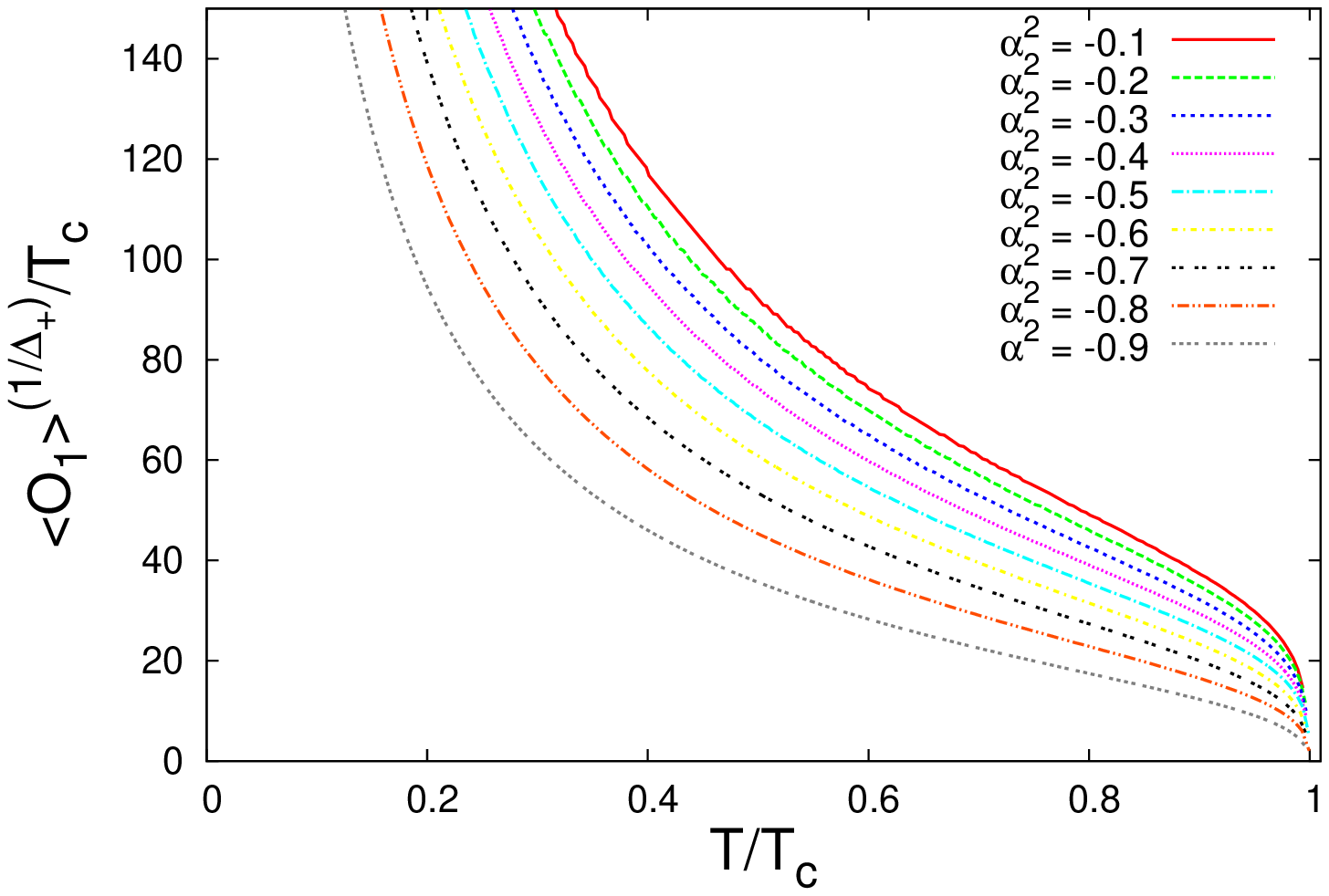}
\includegraphics[height=0.45\textwidth,angle=-90]{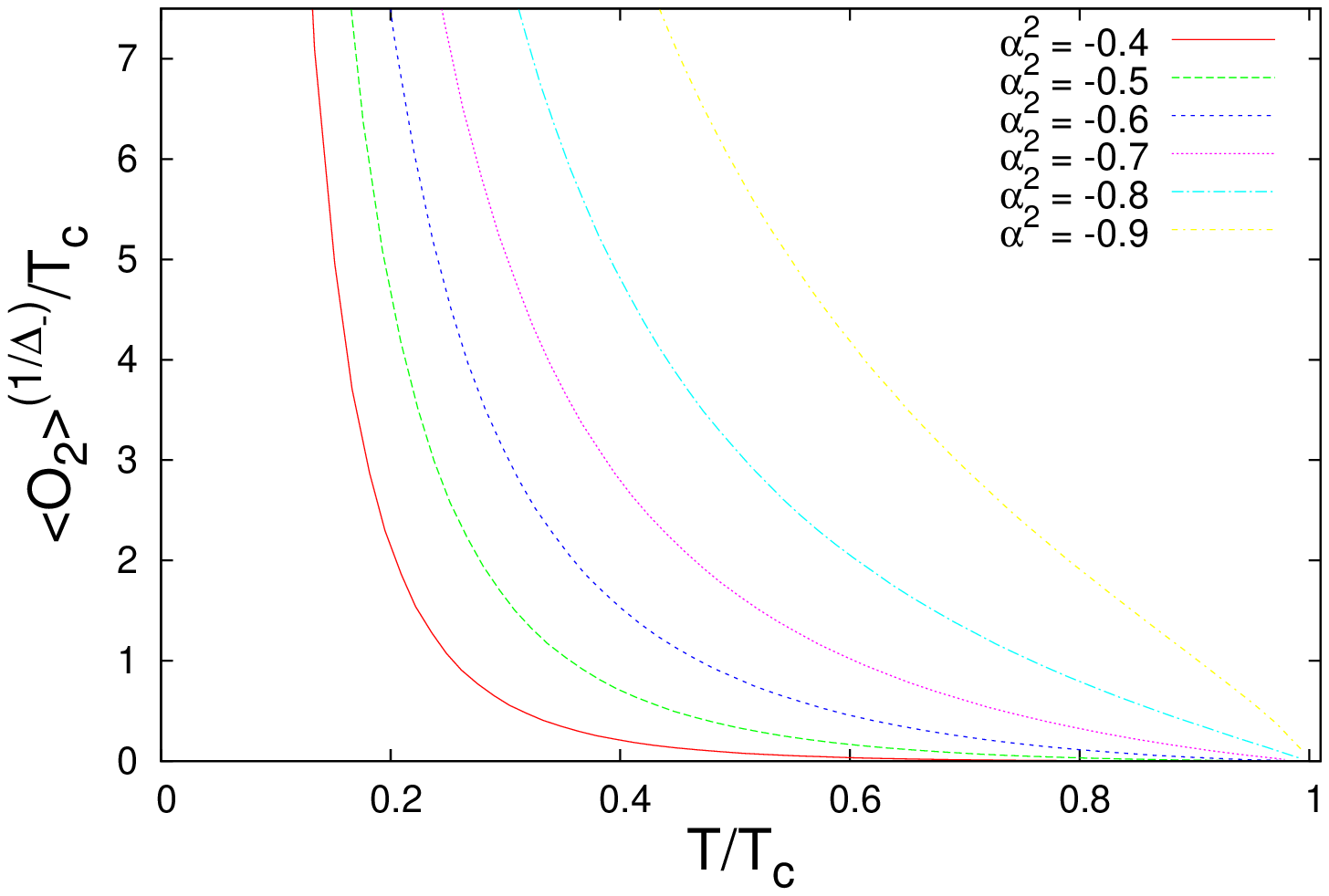}
\caption{Order parameters dependent on $T$ for $z=1$.}
\label{graf_cn_psi_12_T_z_1}
\end{figure}

\begin{figure}[!h]
\centering
\includegraphics[height=0.45\textwidth,angle=-90]{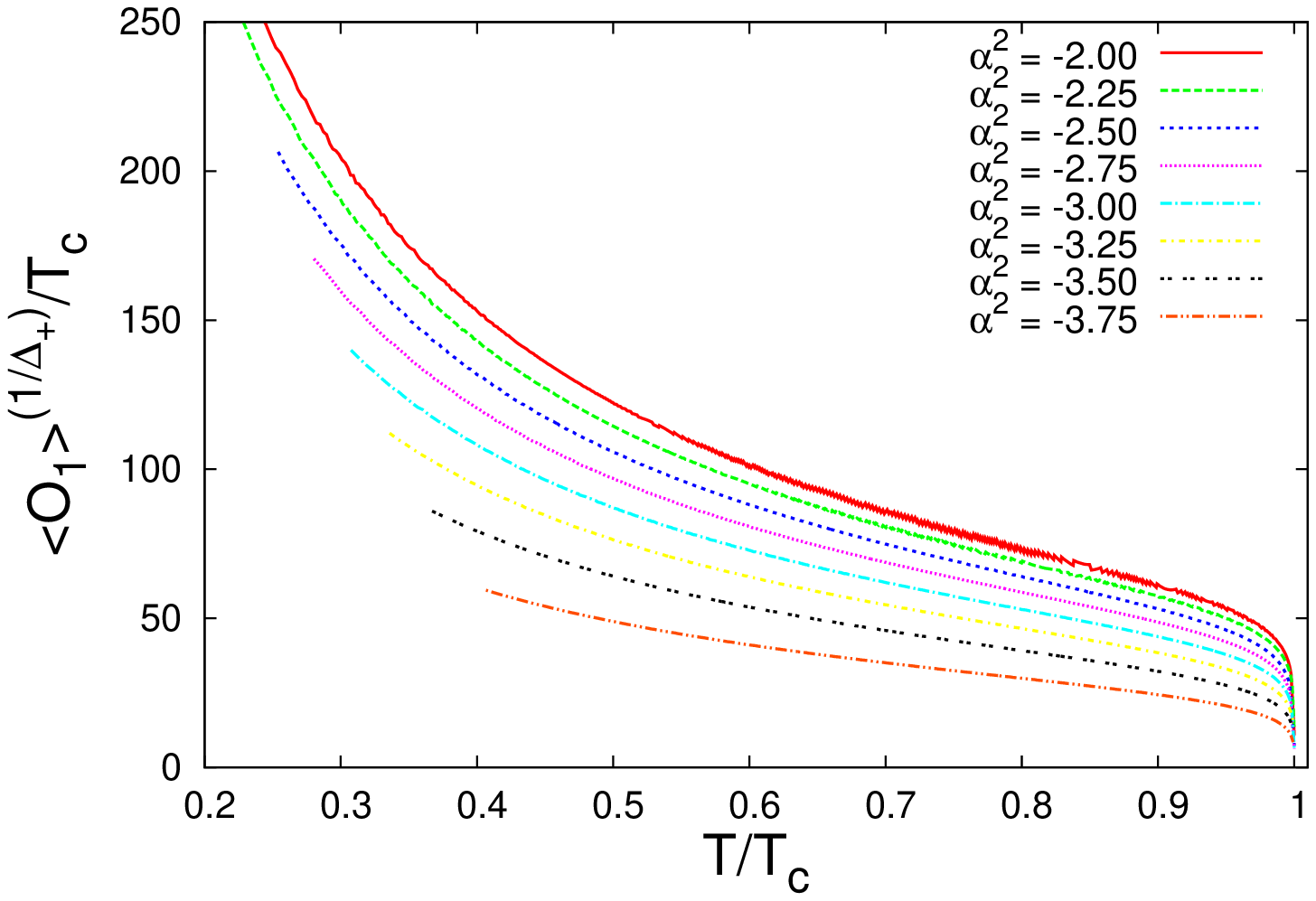}
\includegraphics[height=0.45\textwidth,angle=-90]{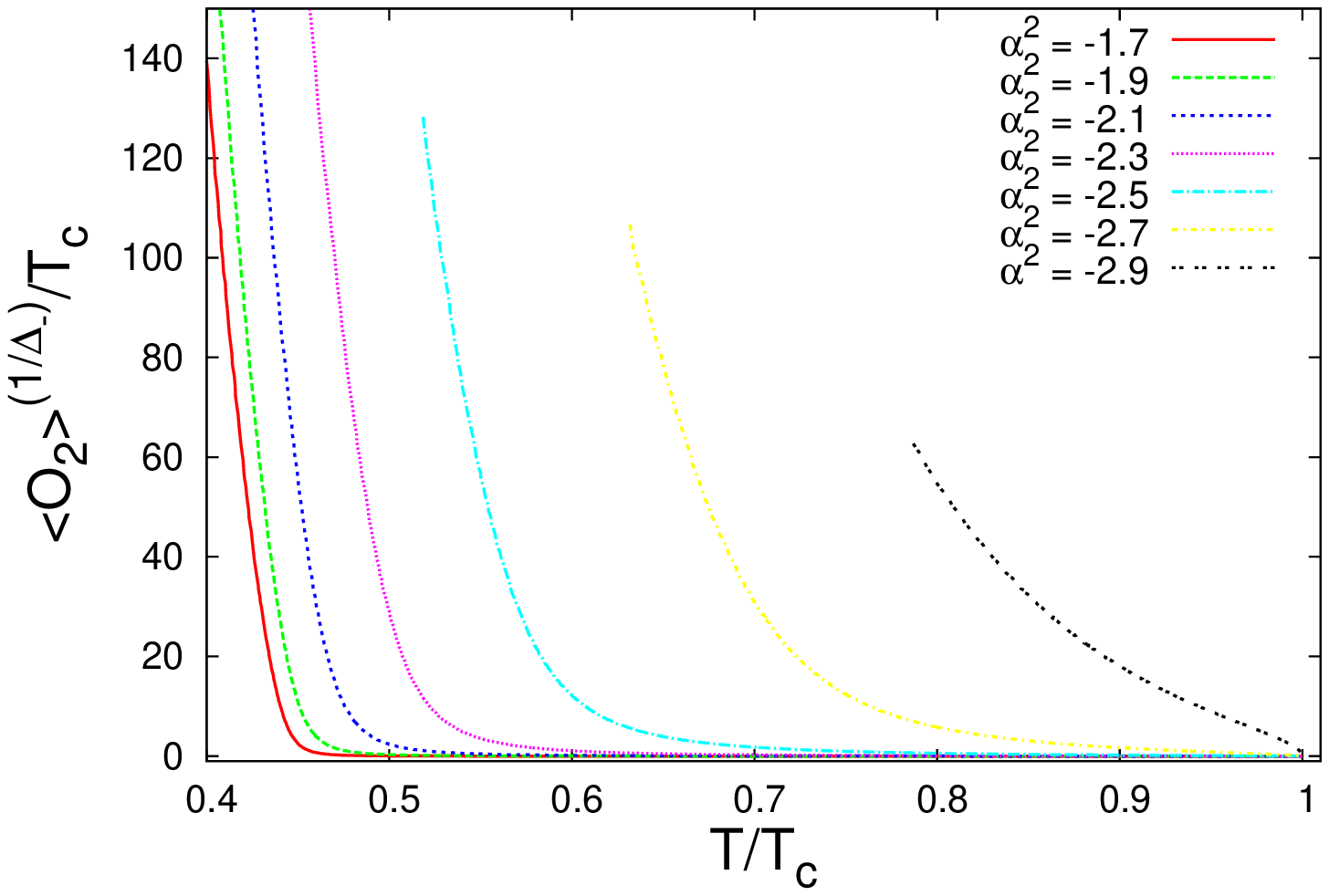}
\caption{Order parameters dependent on $T$ for $z=3$.}
\label{graf_cn_psi_12_T_z_3}
\end{figure}

\begin{figure}[!h]
\centering
\includegraphics[height=0.45\textwidth,angle=-90]{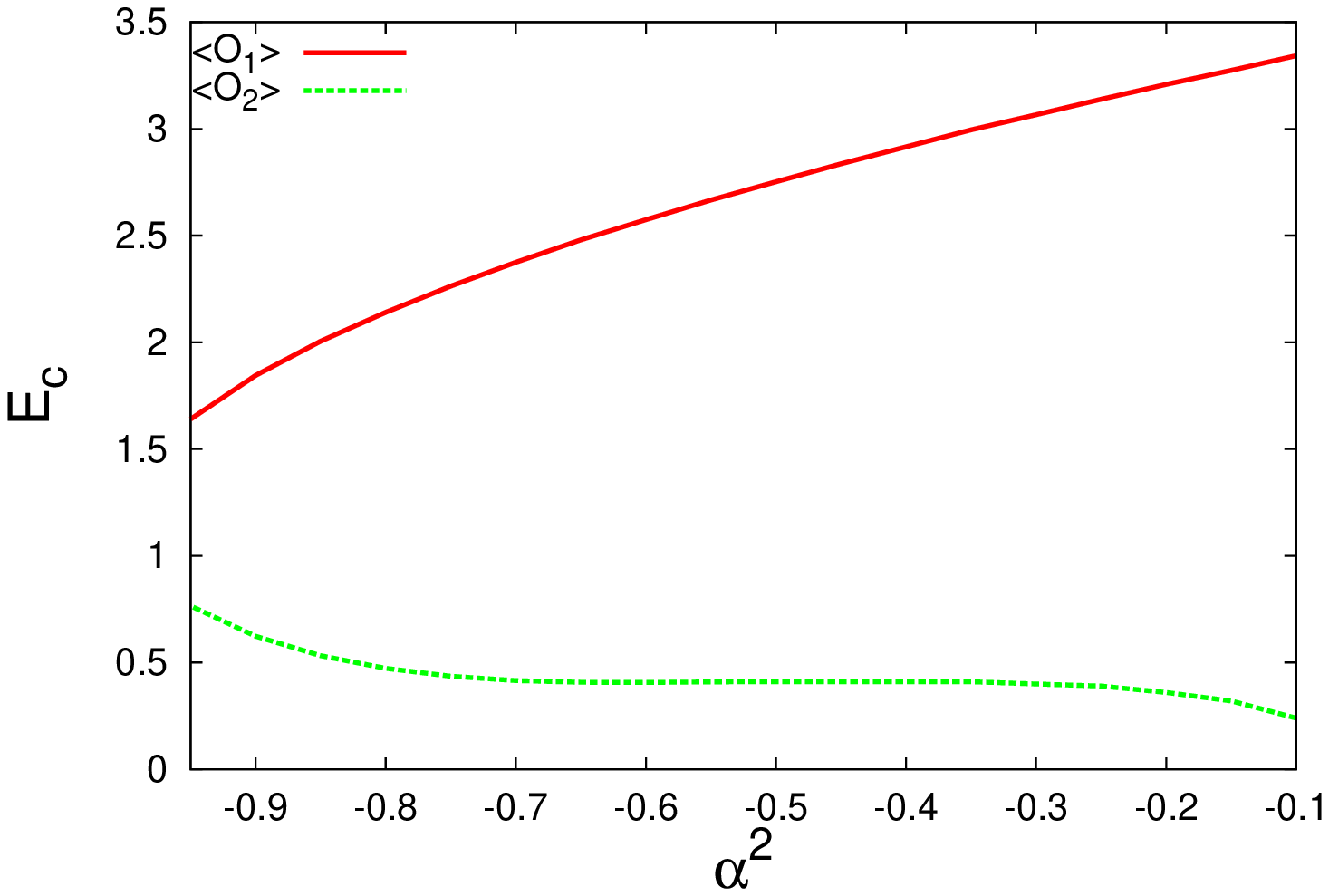}
\includegraphics[height=0.45\textwidth,angle=-90]{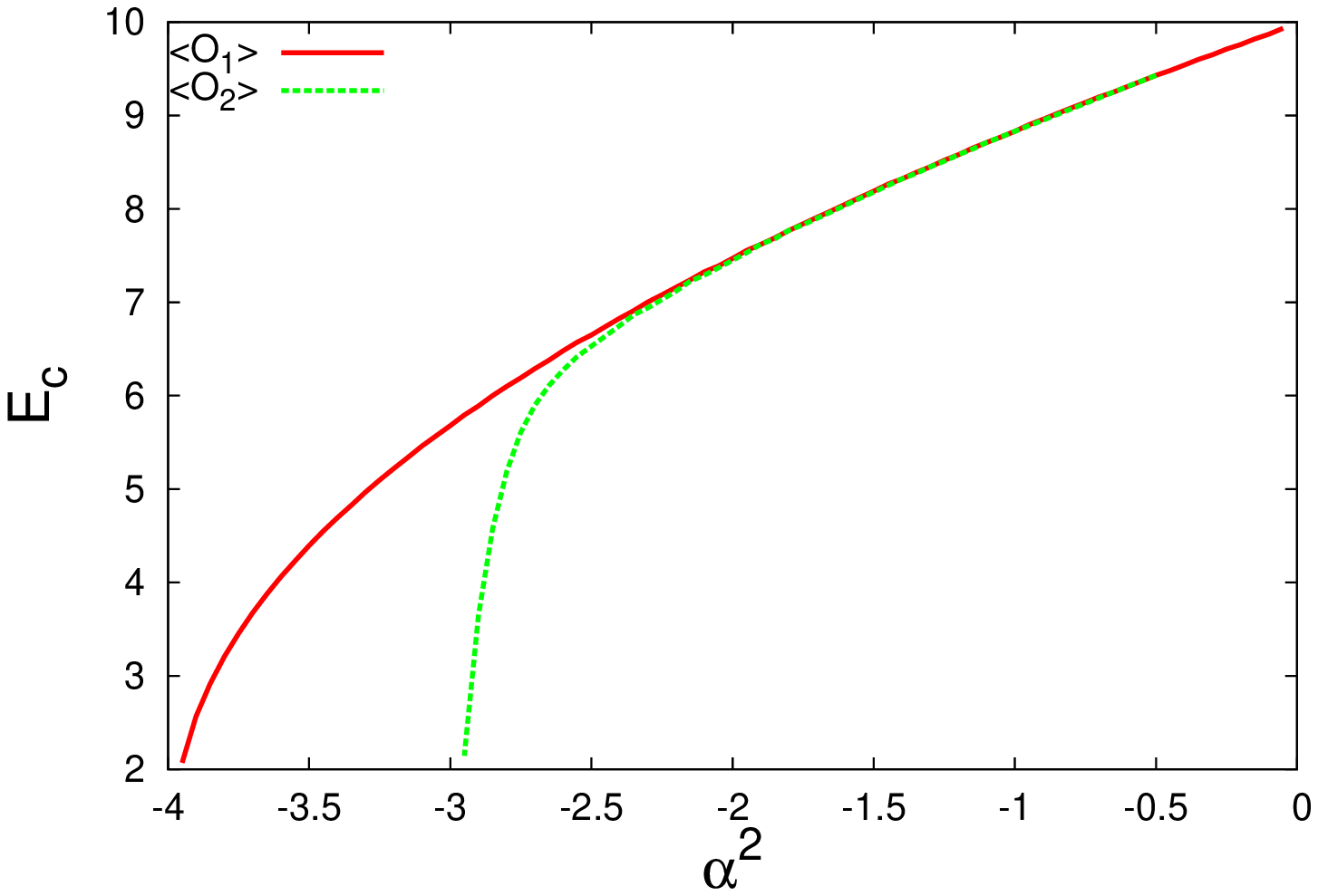}
\caption{Critical electrical field for $z=1$ (left) and $z=3$ (right).}
\label{fit_b_E}
\end{figure}

\begin{figure}[!h]
\centering
\includegraphics[height=0.45\textwidth,angle=-90]{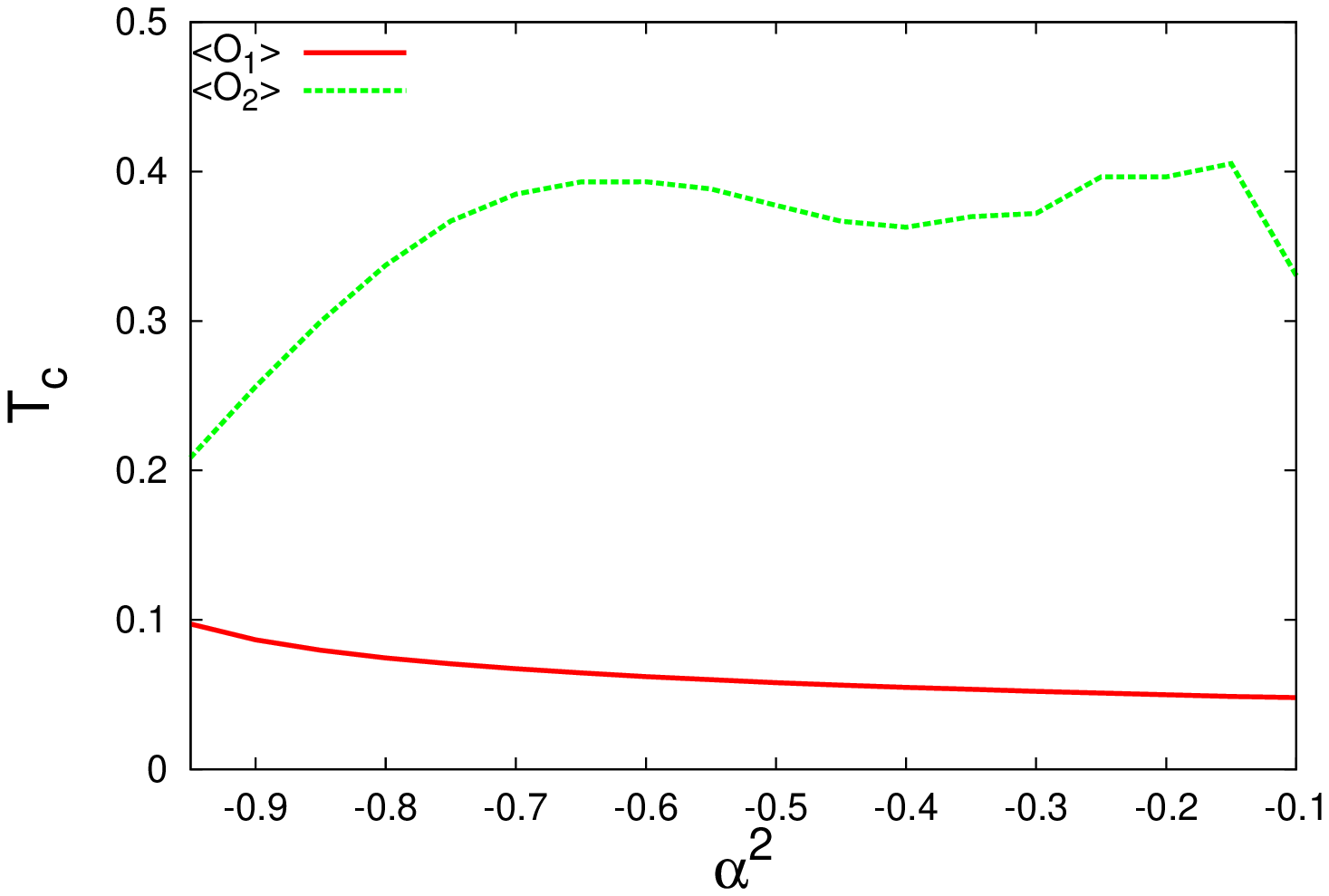}
\includegraphics[height=0.45\textwidth,angle=-90]{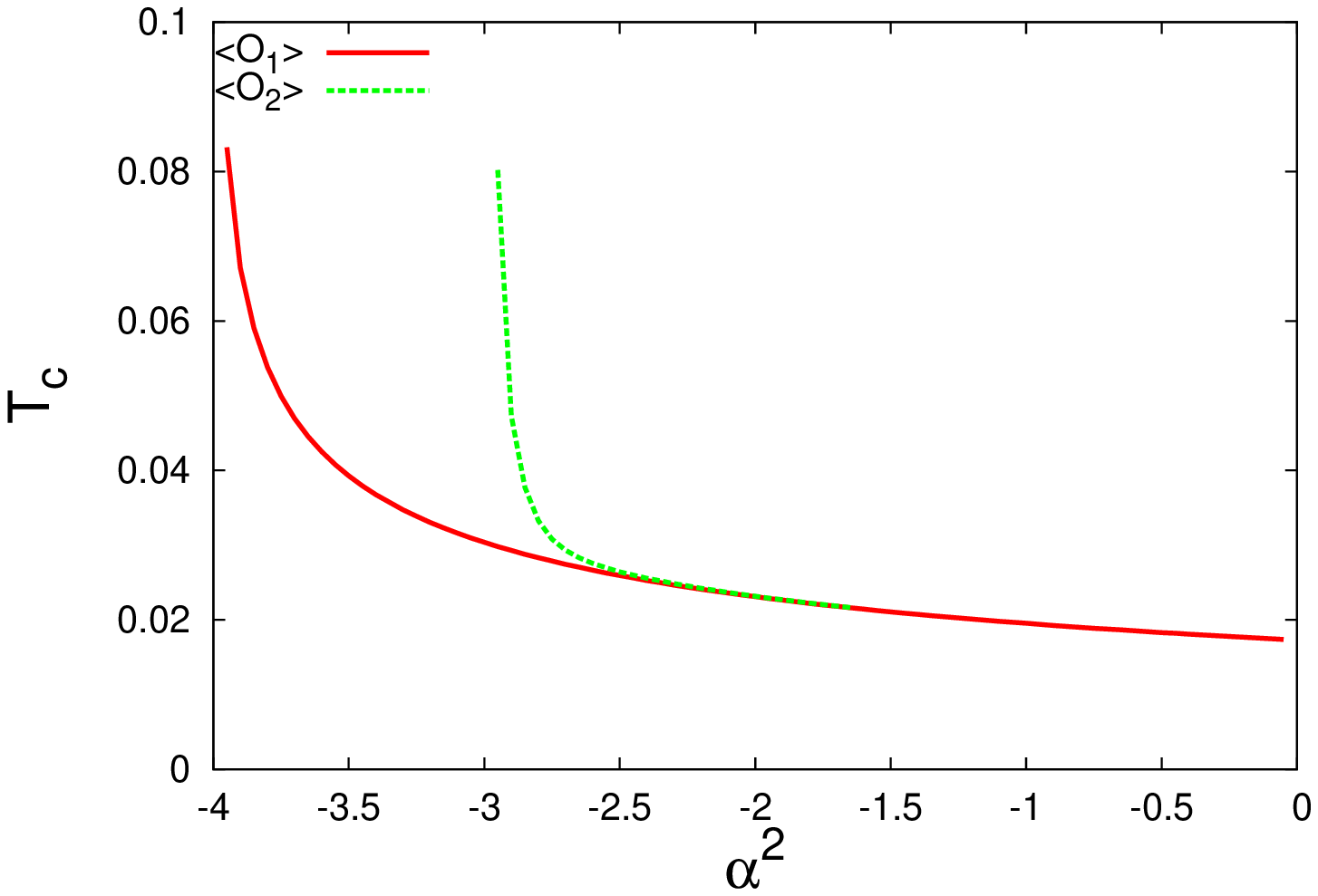}
\caption{Critical temperature for $z=1$ (left) and $z=3$ (right).}
\label{fit_b_T}
\end{figure}

\begin{figure}[!h]
\centering
\includegraphics[height=0.45\textwidth,angle=-90]{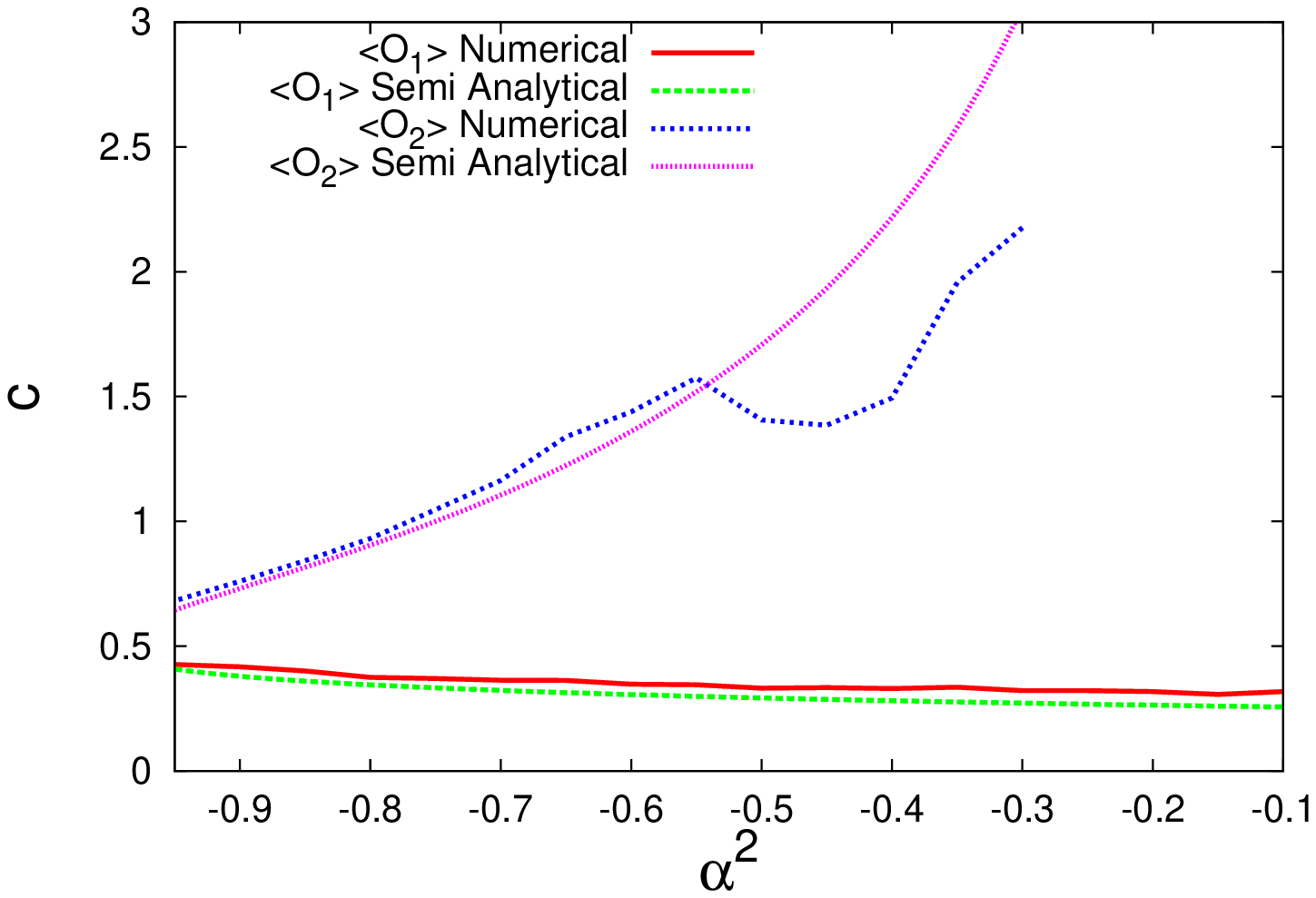}
\includegraphics[height=0.45\textwidth,angle=-90]{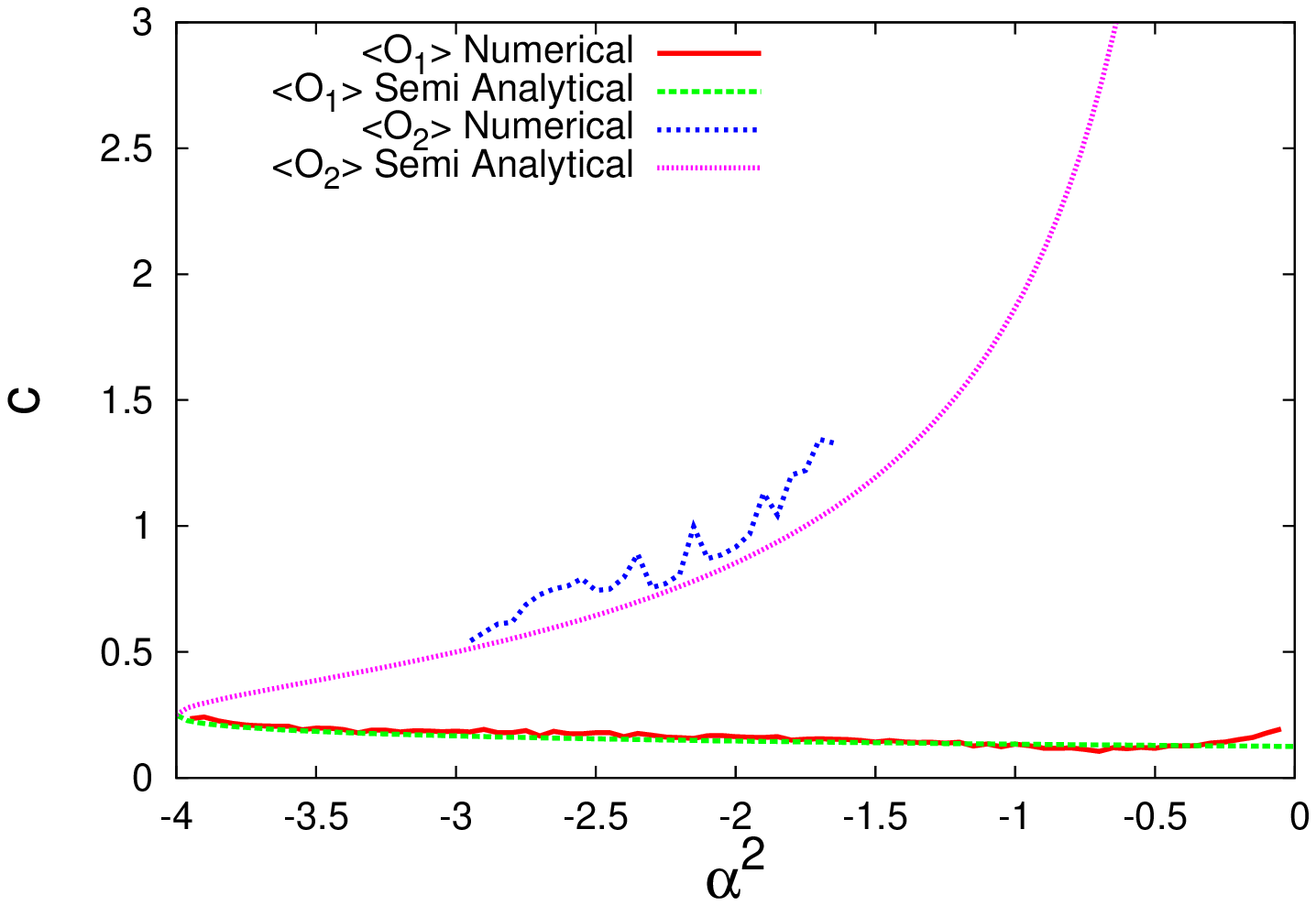}
\caption{Critical exponent dependent on $E_+$ for $z=1$ (left) and $z=3$ (right).}
\label{fit_c_E}
\end{figure}

\begin{figure}[!h]
\centering
\includegraphics[height=0.45\textwidth,angle=-90]{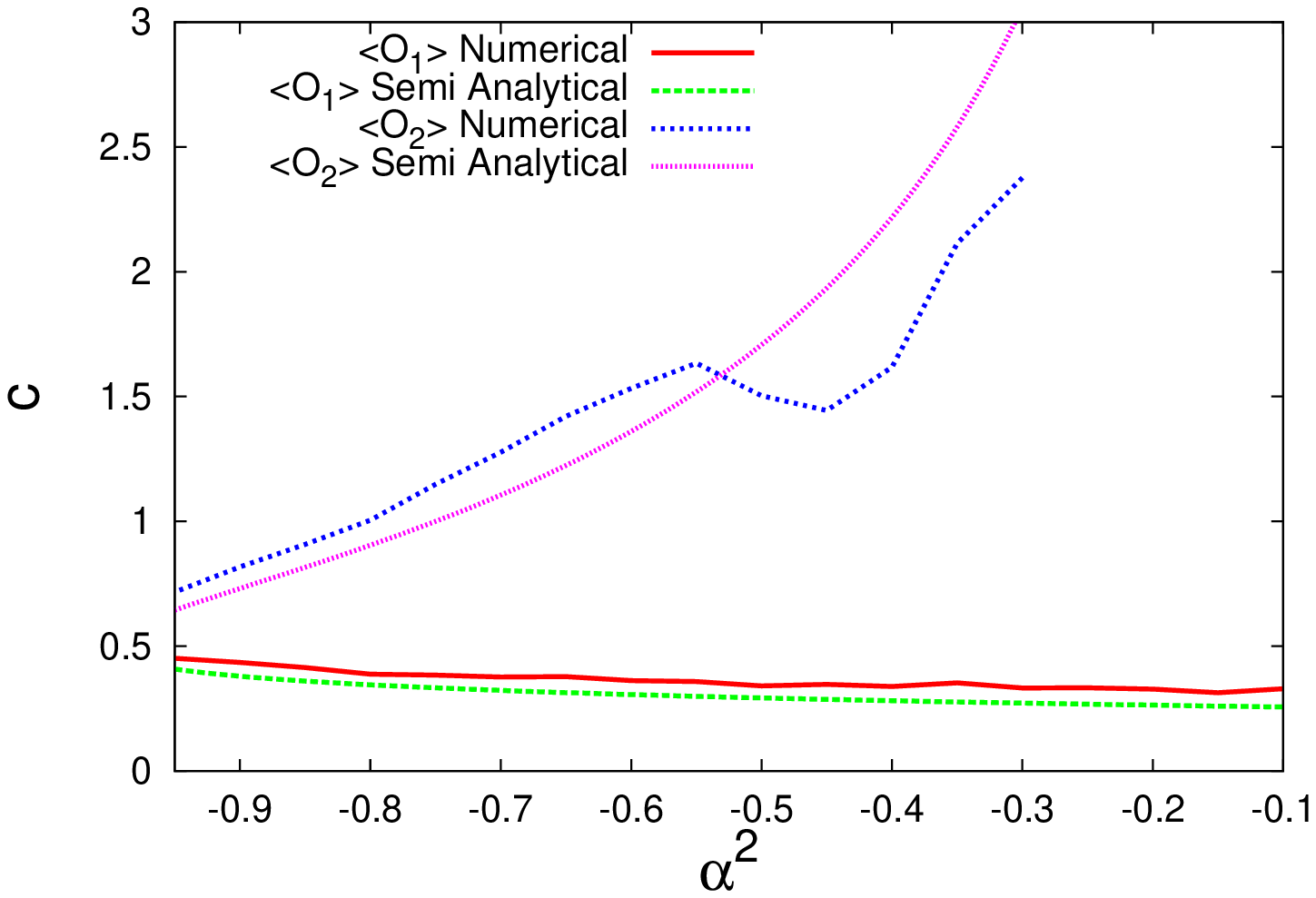}
\includegraphics[height=0.45\textwidth,angle=-90]{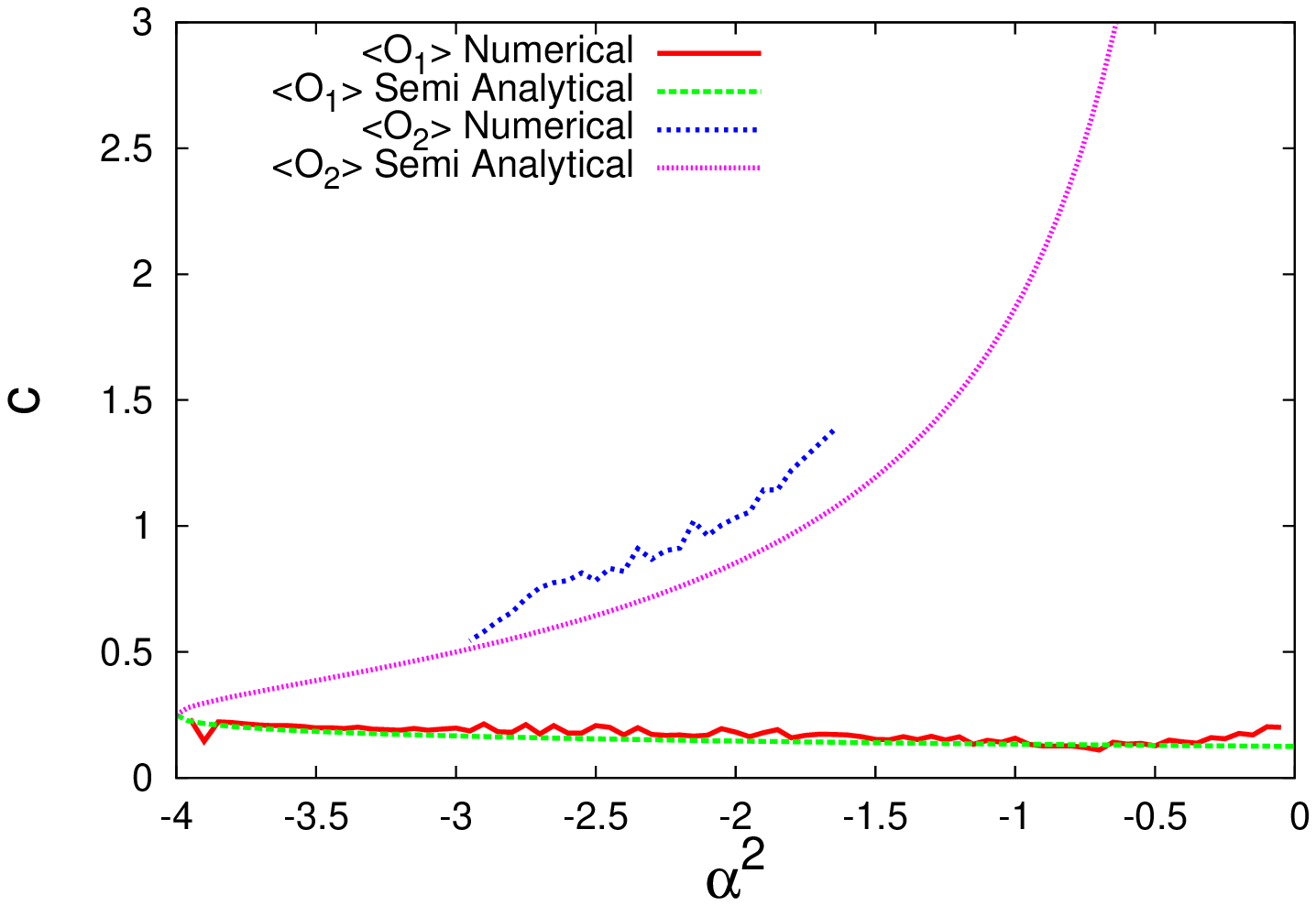}
\caption{Critical exponent dependent on $T$ for $z=1$ (left) and $z=3$ (right).}
\label{fit_c_T}
\end{figure}

The fundamental curve (labelled as $0$) has the lowest critical electric field and highest critical
temperature. We fit the behaviour of the fundamental curves as $y = a (b-x)^c$, where $a$ is not
important, $b$ is the critical electrical field $E_c$ in case of dependence on $E_+$ or the critical
temperature $T_c$ in case of dependence on $T$, and $c$ is the critical exponent. The order parameters are shown in
figures~\ref{graf_cn_psi_12_E_z_1}, \ref{graf_cn_psi_12_E_z_3}, \ref{graf_cn_psi_12_T_z_1},
and~\ref{graf_cn_psi_12_T_z_3} while the fitted parameters $b$ and $c$ are shown in figures~\ref{fit_b_E},
\ref{fit_b_T}, \ref{fit_c_E} and~\ref{fit_c_T}.

We notice that for $\langle O_2 \rangle$ and $z=3$ the critical electrical field goes to zero as $\alpha^2$ reaches
$-3$. For $-4<\alpha^2<-3$, there are no fundamental curves, since the first occurrence of a sign change
in $C_1$ corresponds to boundary conditions for which $\Psi(u)$ changes sign once. As seen in
figure~\ref{delta_limit}, this range of $\alpha^2$ should have been excluded, because it is assumed
that $\Psi(u)$ decays faster than $u$ in order to obtain eq. (\ref{phi_z}). However, we observe
that this asymptotic expressions fit the data derived by Runge-Kutta method and all results for
$\langle O_2 \rangle$ are consistent with the results for $\langle O_1 \rangle$, for which $\Delta_+$
is always bigger than one. The same reasoning is valid for $z=1$. All values of $\alpha^2$ should have been excluded
for $\langle O_2 \rangle$, but eq. (\ref{phi_z}) fits the numerical data even in this case and all results are
consistent with $\langle O_1 \rangle$.

\newpage
\section{Further Remarks}\label{sec-remarks}

The holographic description of a $(1+1)$ dimensional field theory with Lifshitz symmetry displaying
a second order phase transition is presented. This result might imply a contradiction with the Coleman-Mermin-Wagner theorem \cite{coleman,mermin,hohenberg}. However, for a  BTZ  spacetime (namely our $z=1$ case) it has been shown \cite{anninoshartiqbal} that the theorem is evaded by means of a Berezinskii-Kosterlitz-Thouless phase transition
\cite{berezinskii,kosterlitz-thouless} as it has been usual in relativistic two dimensional field theory with mass generation \cite{abw}. For $z=3$ there is a further break of space-time symmetry and a possible prohibition of a phase transition is further removed. In this case there is no ground for any version of the Coleman-Mermin-Wagner theorem. In fact,
there is no global symmetry breaking. Thus, not even a Berezinskii-Kosterlitz-Thouless mechanism is
envisaged, leaving the system free to have a phase transition of the kind found in the present paper.

Some similarities between both cases, $z=1$ and $3$ are pointed out. The fact that there is a phase transition in terms either of a critical electric field or a temperature is very much the same, even the dependence of the order parameters
on the temperature is hardly seen to display any difference, thus indicating that the mechanism of obtaining the phase
transition is very similar in both cases. This may sound a bit surprising, since in the real world superconductivity
and other thermodynamical properties depend a lot on details of the system, while here we have a too robust result,
always similar to the mean field result, independent even on the dimensionality of the system.

The order parameters grow unbounded as $T$ goes to zero. In~\cite{3Hs}, it is argued that this behaviour
indicates that we cannot assume no backreaction for small values of $T$. According to~\cite{herzoga},
our results also suggest strong pairing interactions.  Indeed, the larger value of $\langle O\rangle$
when $T\to 0$ is expected for a strongly interacting field theory. Thus, being a strongly coupled system,
backreaction must be considered, which does not mean that the order parameters do not diverge at
small temperatures. This correction will be studied in future works.

For the conductivity, we tried to solve a differential equation for the field $A_\phi$, which is coupled to
eqs.~\eqref{phi_semT} and~\eqref{psi_semT} 
and fitted an asymptotic behavior similar to eq.~\eqref{phi_z}
to find $\langle J_\mu \rangle$ and $A_\phi^{(0)}$, which we used to find the conductivity $\sigma(\omega)$.
In the results we found peaks consistent with resonances, but we could not explain what caused these resonances.
If we smoothed the data with techniques such as plotting a Bezier curve from the data we could get a behavior consistent with~\cite{3Hs},
but this approach seemed too artificial. Therefore we decided to remove our conductivity analysis to focus on the phase transition.

A possible modification in these results could be obtained considering a Chern-Simons term in the action since that it introduces a new coupling between $A_{t}$ and $A_{\phi}$. In fact, in other works \cite{kraus,clement} the presence of Chern-Simons term solves some problems in $(2+1)$ spacetimes. We will deal with this term in a future work.

%------------------------------------------------------------%

\begin{acknowledgements}
We would like to thank Bin Wang and Eleftherios Papantonopoulos for useful discussions.
This work has been supported by FAPESP, FAPEMIG and CNPq, Brazil.
\end{acknowledgements}

%------------------------------------------------------------%
\section*{References}

\bibliography{references}

\end{document}